\makeatletter
\declare@file@substitution{revtex4-1.cls}{revtex4-2.cls}
\makeatother
\documentclass[twocolumn]{aastex631}
\usepackage{newtxtext,newtxmath}
\usepackage[T1]{fontenc}
\usepackage{CJKutf8}
\usepackage{ae,aecompl}
\usepackage{graphicx}
\usepackage{amsmath}
\usepackage{amssymb}	
\usepackage{float}
\usepackage[]{hyperref}
\newcommand{\ud}{\mathrm{d}}

\usepackage{float}
\usepackage{placeins}

\newcommand{\hamr}{\texttt{H-AMR}}
\usepackage{booktabs}
\received{}
\revised{}
\accepted{}

\submitjournal{ApJ} 
\begin{document}

\title{Novel Polarimetric Analysis of Near Horizon Flaring Episodes in M87* in Millimeter Wavelength} 

\shorttitle{Linear Polarization}
\shortauthors{R. Emami et. al.}

\correspondingauthor{Razieh Emami}
\email{razieh.emami$_{-}$meibody@cfa.harvard.edu}

\author[0000-0002-2791-5011]{Razieh Emami}
\affiliation{Center for Astrophysics $\vert$ Harvard \& Smithsonian, 60 Garden Street, Cambridge, MA 02138, USA}

\author[0000-0003-4475-9345]{Matthew Liska}
\affiliation{Center for Relativistic Astrophysics, Georgia Institute of Technology, Howey Physics Bldg, 837 State Street NW, Atlanta, GA 30332, USA}

\author[0000-0002-2825-3590]{Koushik Chatterjee}
\affiliation{Department of Physics, University of Maryland, College Park, MD, USA}

\author[0000-0003-4056-9982]{Geoffrey C. Bower}
\affiliation{Academia Sinica Institute of Astronomy and Astrophysics, 645 N. A’ohoku Place, Hilo, HI 96720, USA}

\author[0000-0003-2098-170X]{Wystan Benbow}
\affiliation{Center for Astrophysics $\vert$ Harvard \& Smithsonian, 60 Garden Street, Cambridge, MA 02138, USA}

\author[0000-0003-2808-275X]{Douglas Finkbeiner}
\affiliation{Center for Astrophysics $\vert$ Harvard \& Smithsonian, 60 Garden Street, Cambridge, MA 02138, USA}

\author[0000-0002-8635-4242]{Maciek Wielgus}
\affiliation{Instituto de Astrofísica de Andalucía-CSIC, Glorieta de la Astronomía s/n, E-18008 Granada, Spain}

\author[0000-0001-6950-1629]{Lars \ Hernquist}
\affiliation{Center for Astrophysics $\vert$ Harvard \& Smithsonian, 60 Garden Street,  Cambridge, MA 02138, USA}

\author[0000-0003-4284-4167]{Randall Smith}
\affiliation{Center for Astrophysics $\vert$ Harvard \& Smithsonian, 60 Garden Street, Cambridge, MA 02138, USA}

\author[0000-0001-6950-1629]{Grant Tremblay}
\affiliation{Center for Astrophysics $\vert$ Harvard \& Smithsonian, 60 Garden Street,  Cambridge, MA 02138, USA}

\author[0000-0001-5287-0452]{Angelo Ricarte}
\affiliation{Center for Astrophysics $\vert$ Harvard \& Smithsonian, 60 Garden Street, Cambridge, MA 02138, USA}
\affiliation{Black Hole Initiative at Harvard University, 20 Garden Street, Cambridge, MA 02138, USA}

\author[0000-0002-5872-6061]{James F. Steiner}
\affiliation{Center for Astrophysics $\vert$ Harvard \& Smithsonian, 60 Garden Street, Cambridge, MA 02138, USA}

\author[0000-0002-3351-760X]{Avery E. Broderick}
\affiliation{Perimeter Institute for Theoretical Physics, 31 Caroline Street North, Waterloo, ON, N2L 2Y5, Canada}
\affiliation{Department of Physics and Astronomy, University of Waterloo, 200 University Avenue West, Waterloo, ON, N2L 3G1, Canada}

\author[0000-0001-7156-4848]{Saurabh}
\affiliation{Max-Planck-Institut für Radioastronomie, Auf dem Hügel 69, D-53121 Bonn, Germany}

\author[0000-0002-2685-2434]{Jordy Davelaar}
\affiliation{Department of Astrophysical Sciences, Peyton Hall, Princeton University, Princeton, NJ 08544, USA}
\affiliation{NASA Hubble Fellowship Program, Einstein Fellow}

\author[0000-0002-1323-5314]{Josh Grindlay}
\affiliation{Center for Astrophysics $\vert$ Harvard \& Smithsonian, 60 Garden Street, Cambridge, MA 02138, USA}

\author[0000-0001-8593-7692]{Mark Vogelsberger}
\affiliation{Department of Physics, Kavli Institute for Astrophysics and Space Research, Massachusetts Institute of Technology, Cambridge, MA 02139, USA}

\author[0000-0001-6337-6126]{Chi-Kwan Chan}
\affiliation{Department of Astronomy and Steward Observatory, University of Arizona, 933 North Cherry Avenue, Tucson, AZ 85721, USA}
\affiliation{Data Science Institute, University of Arizona, 1230 N. Cherry Ave., Tucson, AZ 85721, USA} 
\affiliation{Program in Applied Mathematics, University of Arizona, 617 N. Santa Rita, Tucson, AZ 85721, USA}

\begin{abstract}
Recent multi-wavelength observations of M87* \citep{2024A&A...692A.140A} revealed a high-energy $\gamma$-ray flare without a corresponding millimeter counterpart. We present a theoretical polarimetric study to evaluate the presence and nature of a potential millimeter flare in M87*, using a suite of general relativistic magnetohydrodynamical simulations with varying black hole (BH) spins and magnetic field configurations. We find that the emergence of a millimeter flare is strongly influenced by both spin and magnetic structure, with limited sensitivity to the electron distribution (thermal vs. non-thermal).
We model the intensity light curve with a damped random walk (DRW) and compare the characteristic timescale ($\tau$) with recent SMA observations, finding that the simulated $\tau$ exceeds observed values by over an order of magnitude. In a flaring case with BH spin a=+0.5, we identify a distinct millimeter flare followed by an order-of-magnitude flux drop. All Stokes parameters show variability near the flare, including a sign reversal in the electric vector position angle. While most $\beta_m$ modes remain stable, the $EB$-correlation phase is highly sensitive to both the flare peak and decay.
We examine polarimetric signatures in photon sub-rings, focusing on modes ns=0 and ns=1. The ns=0 signal closely matches the full image, while ns=1 reveals distinct behaviors, highlighting the potential of space VLBI to isolate sub-ring features. Finally, we analyze the magnetic and velocity field evolution during the flare, finding that magnetic reconnection weakens during the flux decay, and the clockwise velocity flow transitions into an outflow-dominated regime. These results suggest that transient radio variability near flares encodes key information about BH spin and magnetic field structure, offering a novel probe into the physics of active galactic nuclei.   
\end{abstract}


\section{Introduction}
Most recent advances in Very Long Baseline Interferometry (VLBI), particularly with the Event Horizon Telescope (EHT) at sub-millimeter wavelengths, have enabled horizon-scale, high-resolution imaging of supermassive black holes (SMBHs) in nearby galaxies. These observations have provided transformative insights into the structure of SMBHs and the plasma dynamics near their event horizons. The first polarized image of M87* \citep{PaperI,PaperII,PaperIII,PaperIV,PaperV,PaperVI,PaperVII,PaperVIII} revealed a stable morphology during the April 2017 observations, reflecting its long light-crossing timescale of $t_g = GM / c^3 \simeq 9 \, \mathrm{hours}$, where $G$, $M$, and $c$ denote the gravitational constant, black hole mass, and speed of light, respectively. The associated dynamical timescale, defined as a Keplerian period at the innermost stable circular orbit, is of the order of a week or a month, depending on the BH spin value. By contrast, Sagittarius A* (Sgr~A*) exhibits significant and rapid structural variability due to much shorter timescales involved, differing by a factor of $\sim 1400$ \citep{2022ApJ...930L..12A,2022ApJ...930L..13A,2022ApJ...930L..14A,2022ApJ...930L..15A,2022ApJ...930L..16A,2022ApJ...930L..17A,2022ApJ...930L..18F,2022ApJ...930L..19W,2022ApJ...930L..20G,2022ApJ...930L..21B}. The extended dynamical timescale of M87* suggests that multi-year analyses of EHT data could provide valuable insights into the nature and sources of its structural variability \citep{2020ApJ...901...67W, 2025A&A...693A.265E}.

Recent observational studies have extensively explored the time-variability of M87* across multiple frequencies. VLBI observations have provided compelling evidence of time-variability in M87* (see, e.g., \citealt{2007ApJ...660..200L,2016Galax...4...46W,2016ApJ...817..131H,2020ApJ...901...67W}). \citet{2020MNRAS.493.5606J} investigated the origin of jet variability in M87 using shearing synchrotron spot models for both black hole-driven and wind-driven jet scenarios. Their analysis demonstrated that black hole-driven jets generate short-lived, highly variable light curve features, while wind-driven jets produce longer-lived, more stable features. Time-variability in the X-ray band has also been extensively studied \citep{1997MNRAS.284L..21H,2021ApJ...919..110I}. \citet{2021ApJ...919..110I} analyzed archival data from Chandra, NuSTAR, and Suzaku, confirming intraday variability in the X-ray regime. They attributed this variability to particle acceleration occurring within both the core and the jet of M87. In the optical and UV bands, \citet{2009ApJ...699..305H,2011ApJ...743..119P} utilized HST-1 data to investigate polarization and spectral variability in the M87 jet, linking the observed variability to enhanced particle acceleration at shock fronts. Additionally, \citet{2003ApJ...599L..65P,2011ApJ...743..119P} analyzed HST and Chandra observations, reporting month-long timescales for optical variability.
In the TeV band, multiple observatories report variability from M87.
Day-scale variability is observed by the HESS (see, e.g.,  \cite{2006Sci...314.1424A}), MAGIC (see, e.g., \cite{2008ApJ...685L..23A}) and VERITAS collaborations \cite{2009Sci...325..444A}. The HESS observations in 2005 were $\sim$10 times faster than observed in any other wave band.
This implied a compact emission region similar in size to the Schwarzschild radius of the central black hole, with the two most likely regions being the unresolved nucleus of M87 and the HST-1 knot. The VERITAS gamma-ray flares in 2008 were contemporaneous with a strong increase in the radio flux from the nucleus, implying particle acceleration to very high energies very near the central black hole. The 2008 flares were found to be unlikely to originate from the HST-1 knot.  The most rapid TeV flares were seen from M87 using VERITAS in 2010 \cite{2012ApJ...746..141A} and further constrain the emission region size.  Here the trailing edge of a flare had an exponential flux decay time of 0.90 (+0.22, -0.15) days, and the shortest exponential rise time was 2.87 (+1.65, -0.99) days. The TeV observations have long been coordinated with multi-wavelength (MWL) observing campaigns. No unique or
common MWL signature is seen for any of the three aforementioned
TeV flare episodes \cite{2012ApJ...746..151A}, and extensive
MWL campaigns have yet to pinpoint the precise location where
particle acceleration is taking place (see, e.g., \cite{2012ApJ...746..151A}). Recently, these extensive MWL
campaigns regularly include EHT observations, and these
campaigns have covered both a low-TeV flux 
state  \cite{2021ApJ...911L..11E} 
and a minor TeV flare \cite{2024A&A...692A.140A}.

Theoretically, numerous studies have explored the polarimetric properties of M87* using a variety of approaches, including semi-analytic models and GRMHD simulations \citep[see, e.g.,][and references therein]{2022ApJ...931...25T,2022arXiv220513696Z,2022ApJ...929...49P,2021ApJ...922..180P,2020ApJ...896...30A,2021ApJ...923..272E, 2021MNRAS.505..523R, 2023ApJ...950...38E, 2023ApJ...955....6E, 2023Galax..11...11E}. However, most of these analyses have focused on static or time-averaged properties, with considerably less attention paid to the horizon-scale time variability of the hot plasma orbiting the SMBH. Given that GRMHD simulations are inherently scale-free, extended analyses over longer  timescales could reveal signatures of time variability in simulations of M87*.

Several intriguing quantities can be studied to understand structural time-variability. In this context, recent advancements in algorithms originally developed to analyze the structural variability of Sgr~A* offer significant potential. For instance, \citet{2009ApJ...695...59D} introduced non-imaging algorithms that leverage interferometric closure quantities in total intensity to detect structural periodicity and time-variability in VLBI observational data of Sgr~A*. This method was later expanded by \citet{2009ApJ...706.1353F} to incorporate full polarimetric data from VLBI observations, uncovering variability across timescales ranging from hours to years (see, e.g., \citealt{2005ApJ...618L..29B,2006ApJ...646L.111M,2006ApJ...640..308M,2007ApJ...654L..57M} and references therein). Rapid variability on timescales as short as sub-minute \citep{2022ApJ...930L..19W} and polarimetric variability on dynamical timescales potentially associated with orbital motion \citep{Wielgus_hotspot,Yfantis2024} were identified in the Sgr~A* mm wavelength light curves.

The most recent broadband, multi-wavelength observations of M87 \citep{2024A&A...692A.140A}, conducted during the EHT observational campaign in 2018, uncovered a high-energy flaring episode in $\gamma-$rays. Inspired by this groundbreaking discovery, in this manuscript, we undertake a comprehensive investigation into the time-variability of M87* in the vicinity of a flaring event, characterized by the global maximum in flux intensity. Employing a suite of state-of-the-art GRMHD simulations with varying BH spins, we systematically analyze the polarimetric signatures of flaring events. Our study focuses on a MAD simulation for a BH spin of  a=+0.5, enabling detailed exploration of the interplay between magnetic fields, spin, and plasma dynamics during a flare. We rigorously identify parameters that exhibit pronounced variability and those that remain stable, leveraging these insights to propose a diagnostic framework for characterizing flaring events.

The paper is organized as follows. Secs. \ref{hamr}  and \ref{imaging} provide an overview of \hamr{} simulations as well as the BH images. In Sec.  4 we identify flaring events in the long-term light curve of M87 and present the black hole images and emission properties in the vicinity of a flaring event. Sec. \ref{Polarimetric-Flare-Image} focuses on the polarimetric analysis of M87* in the image space near the flux flare. In Sec. \ref{Polarimetric-Flare-Image-Visibility-space}, we extend the polarimetric analysis to the visibility space. Sec. \ref{sub-ring} examines the signatures of the flaring event in the photon ring. Secs. \ref{Magnetic-field-flare} and \ref{Velocity-field-flare} investigate the structure of the magnetic field and the velocity field, respectively, in the vicinity of the flaring event. Finally, we present our conclusions in Sec. \ref{Conclusion} and outline potential directions for future research in Sec. \ref{future}.

\begin{figure*}[t!]
\center
\includegraphics[width=1.01\textwidth]{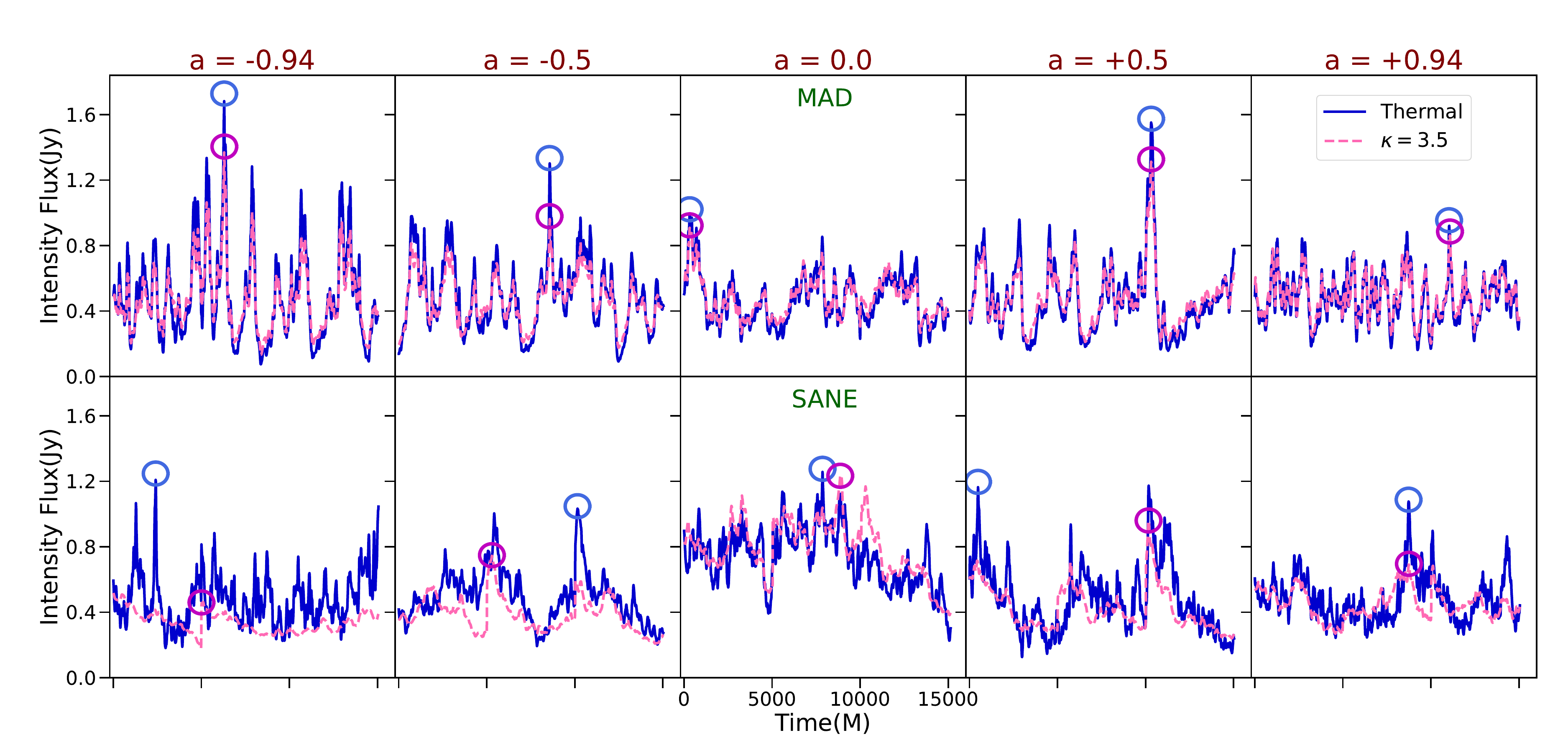}
\caption{The light curve of M87* is presented for thermal and non-thermal kappa models (\(\kappa = 3.5\)) within MAD (top row) and SANE (bottom row) simulations. In each row, from left to right, we increase the BH spin across the sequence a = (-0.94, -0.5, 0.0, +0.5, +0.94). 
In both thermal (blue) and non-thermal (pink) cases, we identify the moments of global maximum flux, referred to as the "BH Flare," denoted by cyan and yellow circles, respectively. Notably, in MAD simulations, the flaring moments align remarkably well between the thermal and non-thermal models. However, in SANE simulations, the maximum flux moments do not necessarily coincide between these two models, underscoring the key role of turbulence in shifting the flux flare in SANE simulations.}
\label{light-curve-M87}
\end{figure*}

\begin{figure*}[t!]
\center
\includegraphics[width=1.01\textwidth]{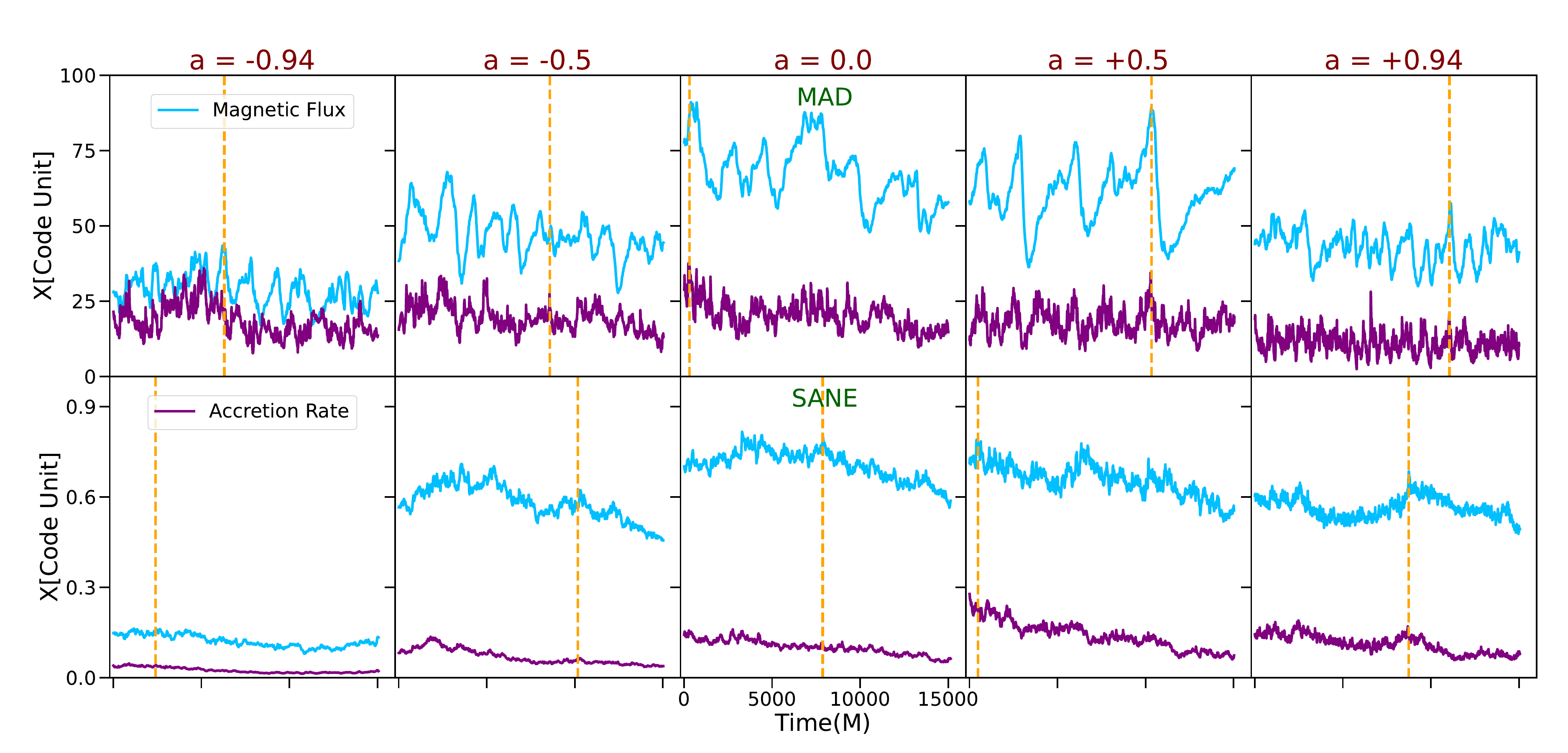}
\caption{The light curve of the magnetic flux (blue) and the accretion rate (purple) is depicted for M87* MAD (top row) and SANE (bottom row) simulations. In each row, from left to right, we increase the BH spin across the sequence a = (-0.94, -0.5, 0.0, +0.5, +0.94). In each panel, the flare time, taken from Figure \ref{light-curve-M87} is shown as a dashed orange line. Both the magnetic flux and the accretion rate are depicted in the code unit.
Notably, in MAD simulations, in the vicinity of the flaring time, the (growing/suppressing) behavior of the magnetic flux and the accretion rate are fairly resembling their corresponding behavior in the light curve. }
\label{light-curve-Magnetic-Flux-Accretion}
\end{figure*}

\begin{figure*}[t!]
\center
\includegraphics[width=1.01\textwidth]{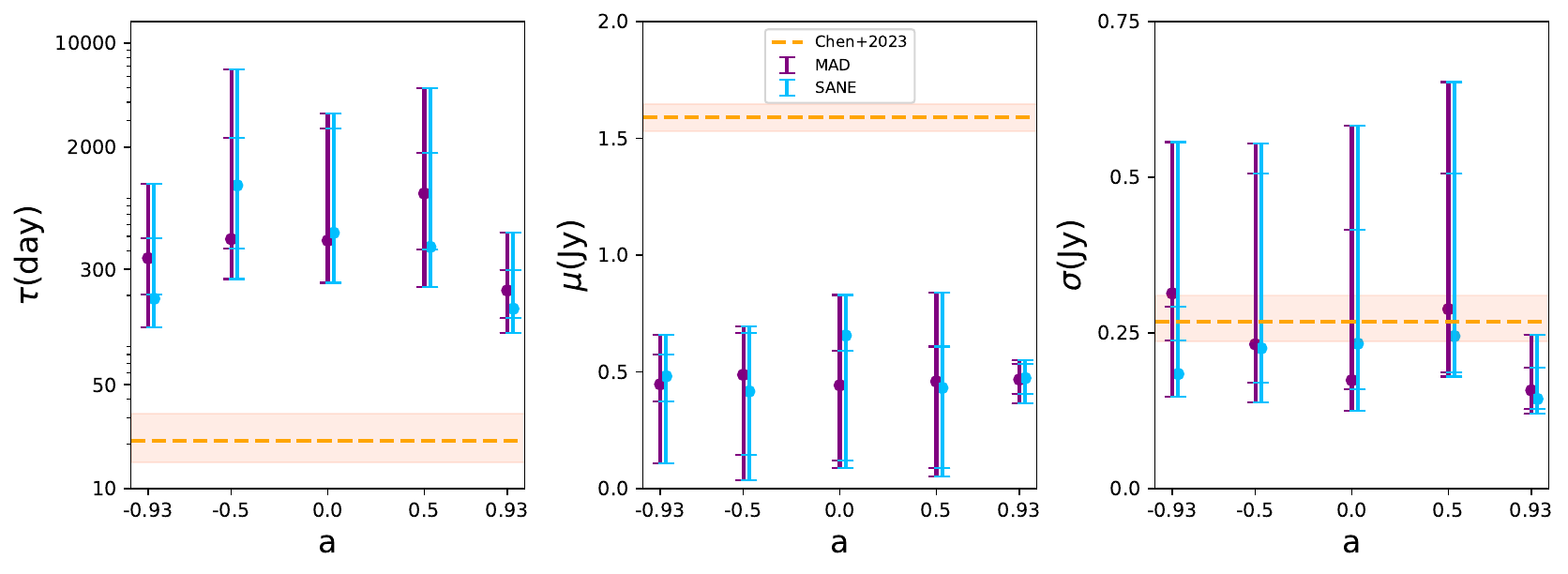}
\caption{Inferred DRW parameters from GRMHD simulations of MAD (blue-line) and SANE (purple-line) with different BH spins. From the left to right we present the de-correlation timescale $\tau$, the mean flux $\mu$, and the standard deviation $\sigma$ of the DRW, respectively. Overlaid in each panel we also depict the SMA observational results from \citep{2023ApJ...951...93C}. Since the SMA observation captures the flux both near the event horizon and in the jet base, both $\tau$ and $\mu$ differ from their theoretical estimations focusing only on the near horizon region.}
\label{DRW-inferred-params}
\end{figure*}
\section{HAMR simulation} \label{hamr}
We utilize \hamr{} \citep{2022ApJS..263...26L} to solve the GRMHD equations in a fixed Kerr spacetime. The simulations are conducted using logarithmic Kerr-Schild coordinates and geometric units, where \(G = c = 1\). Consequently, the length and time scales are normalized to the gravitational radius, \(r_{\rm g} = G M/c^2\), and the black hole light-crossing time, \(t_{\rm g} = GM/c^3\), respectively, where \(M\) represents the black hole mass. These simulations form part of the EHT GRMHD simulation library (\citealt{2022ApJ...930L..16A}, also see \citealt{Chatterjee:2022}). The computational grid is axisymmetric and uniformly distributed in \(\log{r}\), \(\theta\), and \(\varphi\). Boundary conditions include outflowing radial boundary conditions, transmissive polar boundary conditions, and periodic \(\varphi\)-boundary conditions, as described in \citep{Liska:2018}.

For our accretion disk, we adopt a constant angular momentum hydrodynamic equilibrium torus solution \citep{Fishbone:1976}. The gas thermodynamics are governed by the ideal equation of state with a constant adiabatic index. Two distinct magnetic field configurations are considered. The first configuration results in a weakly magnetized, gas-pressure-dominated turbulent flow, commonly referred to as the standard and normal evolution (SANE) mode \citep{2003ApJ...599.1238D, GammieIHARM2003, 2012MNRAS.426.3241N}. The second configuration generates strong vertical magnetic fields capable of disrupting accretion, leading to the magnetically arrested disk (MAD) mode \citep{1974Ap&SS..28...45B, 2003ApJ...592.1042I, 2003PASJ...55L..69N}.

The outer boundaries for the SANE and MAD models are set at \(200M\) and \(1000M\), respectively. The grid resolutions employed are \(240 \times 192 \times 192\) for SANE and \(348 \times 192 \times 192\) for MAD. Both models utilize a range of BH spins, including a = (0.0, \(\pm\)0.5, \(\pm\)0.94). The adiabatic indices for the SANE and MAD tori are chosen as 5/3 and 13/9, respectively. Additionally, density is injected into the funnel region wherever the magnetization exceeds 100 for both SANE and MAD simulations, following the method outlined in \citet{Ressler:2017}.

\section{BH Imaging}  
\label{imaging}  
To generate BH images, we employ the general relativistic radiative transfer (GRRT) algorithm implemented in {\sc ipole} \citep{Moscibrodzka&Gammie2018}. Each image is produced with a field of view (FOV) of 200 $\mu$as and a resolution of $400 \times 400$ pixels. The imaging process incorporates synchrotron emission, self-absorption, Faraday rotation, and Faraday conversion. 

Throughout this analysis, we adopt parameters specific to M87*, with $M = 6.2 \times 10^9 M_{\odot}$, located at a distance of $D = 16.9$ Mpc from an observer on Earth. The mass density is adjusted to match the observed flux at 230 GHz, set to $F_{\nu} = 0.5$ Jy \citep{PaperIV,EHTC_M87_2025}. Following \citet{PaperV}, we assume an inclination angle of 17 degrees for retrograde spins and 163 degrees for prograde spins. The simulated images are subsequently rotated to align with the observed position angle of the M87* forward jet at -72 degrees \citep[e.g.,][]{Walker2018}.  

As the current GRMHD simulations assume identical temperatures for electrons and ions, we implement a post-processing approach to model the electron temperature. In this framework, thermal equilibrium is replaced by a collisionless plasma model, allowing electrons and ions to maintain different temperatures \citep{Shapiro+1976, Narayan+1995}. Following \citet{Monika+2016, PaperV, PaperVIII}, we define the ion-to-electron temperature ratio as:  
\begin{equation}  
\frac{T_i}{T_e} = R_\mathrm{high} \frac{\beta^2}{1+\beta^2} + R_\mathrm{low} \frac{1}{1+\beta^2},  
\end{equation}  
where $\beta$ represents the gas-to-magnetic pressure ratio, and $R_\mathrm{low}$ and $R_\mathrm{high}$ are free parameters. For this study, we constrain these parameters to $R_\mathrm{low} = 1$ and $R_\mathrm{high} = 20$.

\section{Flaring event in light curve of M87*}
\label{Sec:Flare-M87}
Flaring events in BH systems manifest as intensity peaks, often traced through the light curve of intensity. Inspired by this, we conduct a comprehensive study of flaring events in M87* using GRMHD simulations with \hamr{}. Specifically, in 
Sec. \ref{AAA1}, we analyze the light curve to identify the peaks in the intensity flux. In Sec.~\ref{B-field-accretion} we study the time evolution of the magnetic flux as well as the accretion rate. In Sec.~\ref{DRW-fit} we present the damped random walk fitting to the light curve and compare its parameters with SMA observations. 
In Sec.~\ref{Wavelet-trans} we introduce the wavelet transformation to simulate the flux flare for different GRMHD simulations. 
In Sec.~\ref{flare-image}, we examine variations in the BH images near these flaring events. Finally, in Sec.~\ref{Emission-location-BH}, we trace the emission location associated with a flare to explore its spatial origin.

\subsection{Long light curve of M87* } 
\label{AAA1}
In this section, we analyze the long-term light curve of M87*, identifying global intensity peaks. Figure~\ref{light-curve-M87} presents the intensity flux light curve for a series of GRMHD simulations with varying BH spins. These simulations span \(15000M \approx 15\) years of observations. The top row depicts the flux for MAD simulations, while the bottom row shows the flux for SANE simulations. Within each row, the BH spin varies from left to right as a = (-0.94, -0.5, 0.0, +0.5, +0.94). 

Each panel overlays the flux for both thermal (blue lines) and non-thermal models with \(\kappa = 3.5\) (pink lines). The global maximum flux is marked by light-blue circles for the thermal models and magenta circles for the non-thermal models. The results indicate that, in MAD simulations, the location of the flux maximum is consistent between the thermal and non-thermal models, regardless of BH spin. In contrast, SANE simulations often exhibit a temporal shift in the flux maximum between the two models, depending on spin. There are different possible reasons for this behavior. Firstly, for MADs, the bulk of the emission comes from the midplane \citep{2023ApJ...950...38E}, which is thermally dominated, While for SANEs with thermal emission, the jet boundary lights up more and therefore has a significant non-thermal contribution to the flux. On the contrary, SANEs with non-thermal emissions are more disk-dominated than thermal ones. Consequently, because the flux is coming from a larger region, the variability decreases.
The spin dependence is also matched with \citep{2022MNRAS.511.3795N} as the jet shape depends on spin values. Furthermore, we also argue that part of the shift in the intensity peak in SANEs might be associated with the turbulent dominance in SANEs, leading to intensity profiles that do not necessarily align between the thermal and non-thermal cases. Furthermore, since the emission region differs between the thermal and the non-thermal models their intensity peaks are also different. 

\begin{figure*}[t!]
\center
\includegraphics[width=1.01\textwidth]{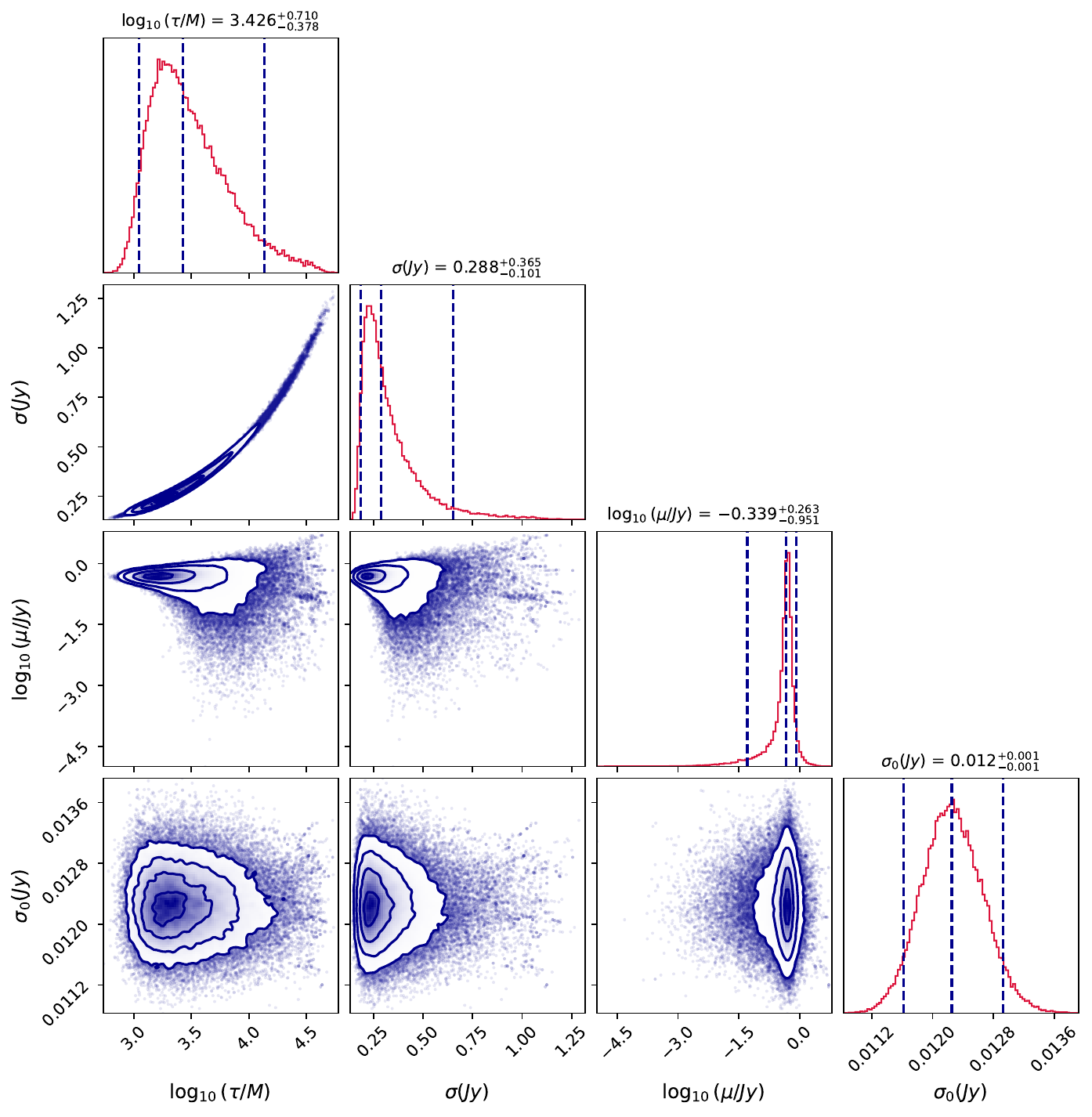}
\caption{The posterior distribution corner plot
for a MAD simulation with a = + 0.5. Depicted for each parameter is the median and the (5, 95) percentiles. }
\label{DRW-MCMC}
\end{figure*}

\begin{figure*}[th!]
\center
\includegraphics[width=1.03\textwidth]{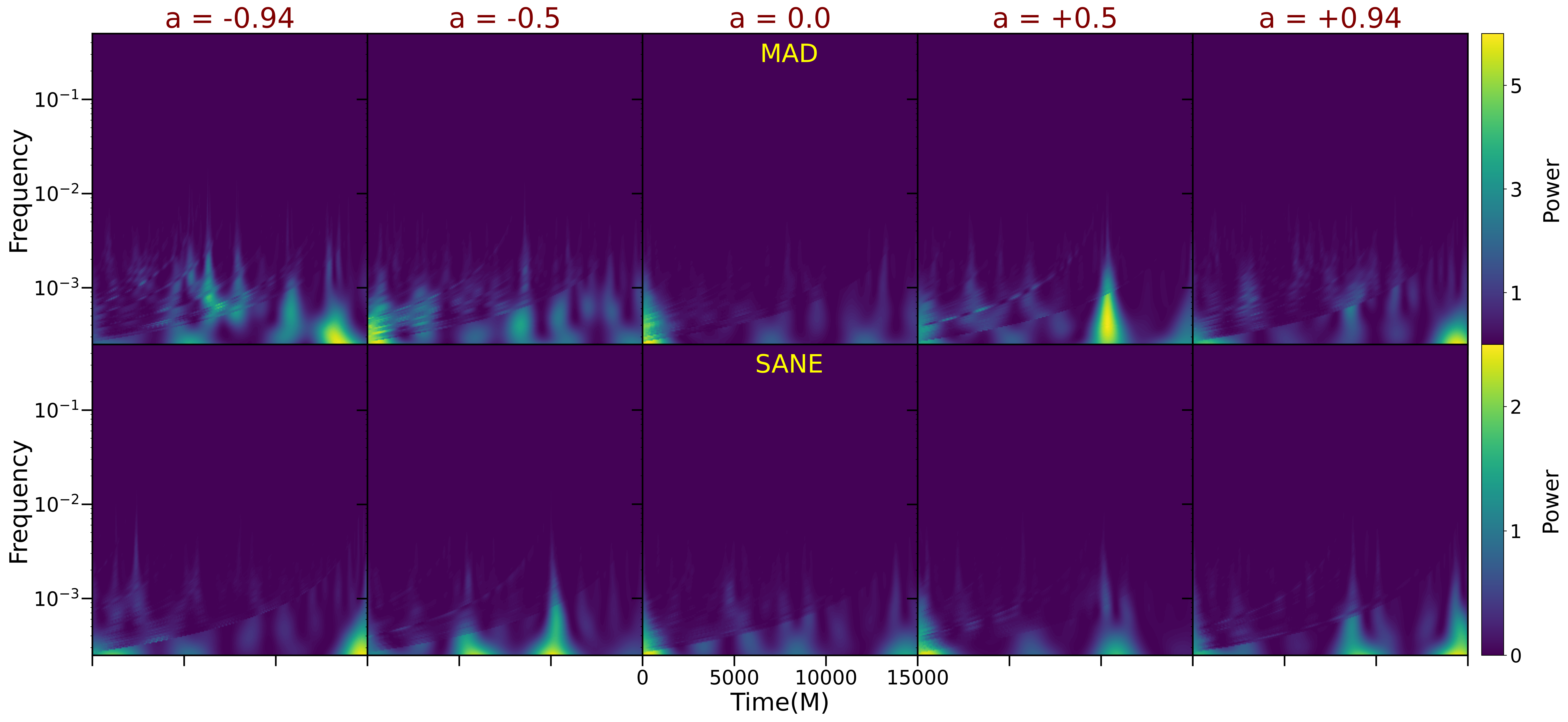}
\caption{Depiction of the wavelet power-spectrum as color contours in the frequency-time plane for the thermal model. It highlights the temporal evolution of frequency components. 
Flux flares appear as regions with higher power. 
}
\label{Wavelet-Transformation}
\end{figure*}

\begin{figure*}
\center
\includegraphics[width=1.0\textwidth]{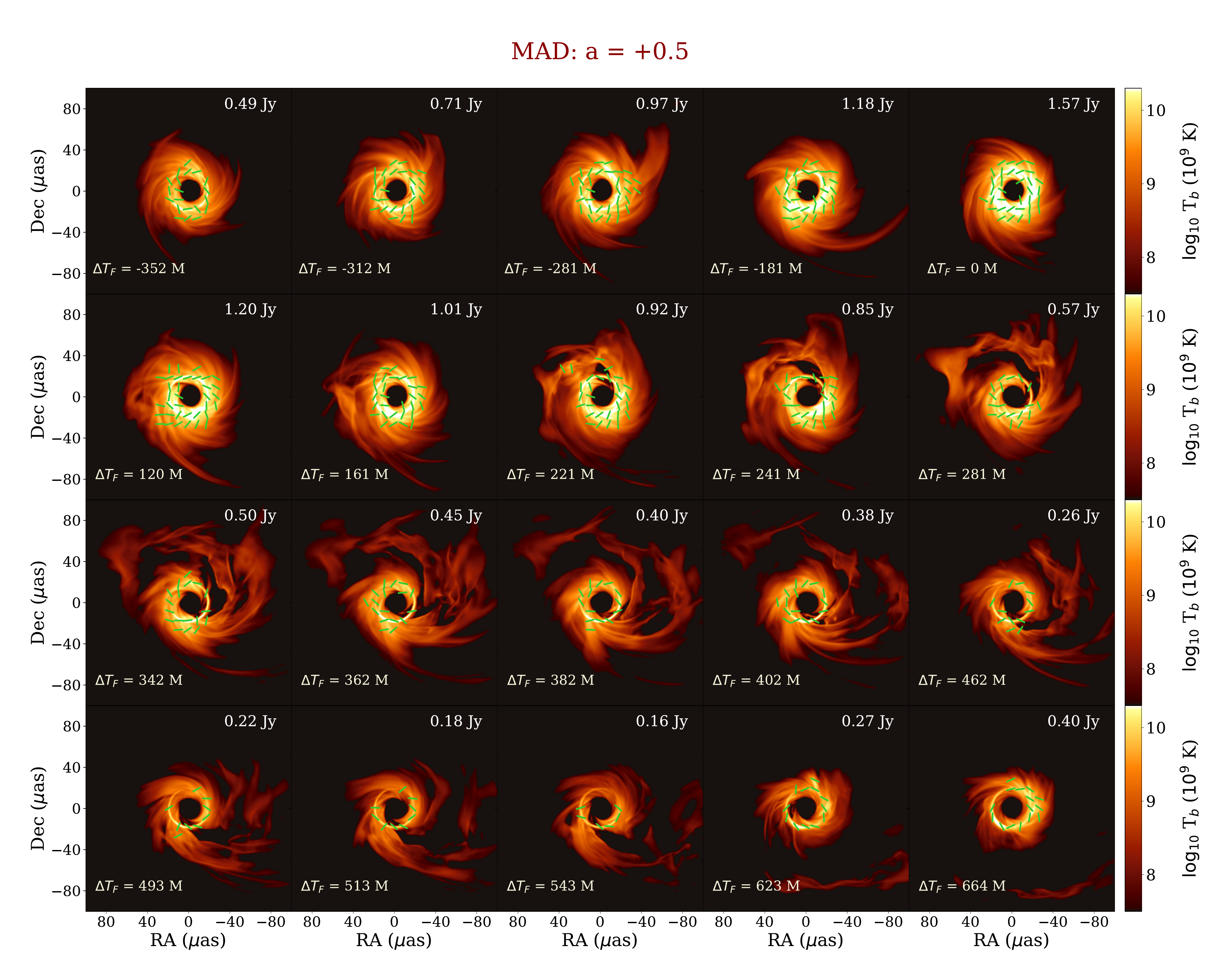}
\caption{BH images in the vicinity of an actual flare for a MAD simulation with $a = +0.5$ spin with a thermal emission. $\Delta T_F$ refers to the time difference from a flare. The flare corresponds to the intensity peak at $I = 1.57$ Jy at $\Delta T_F$ = 0 M. Subsequently, the flux diminishes as it gets pushed out. The end of the flare corresponds to the time with a minimum intensity. After this event, the flux gradually increases.}
\label{Image-Flare-ap5}
\end{figure*}

\begin{figure*}
\center
\includegraphics[width=0.99\textwidth]{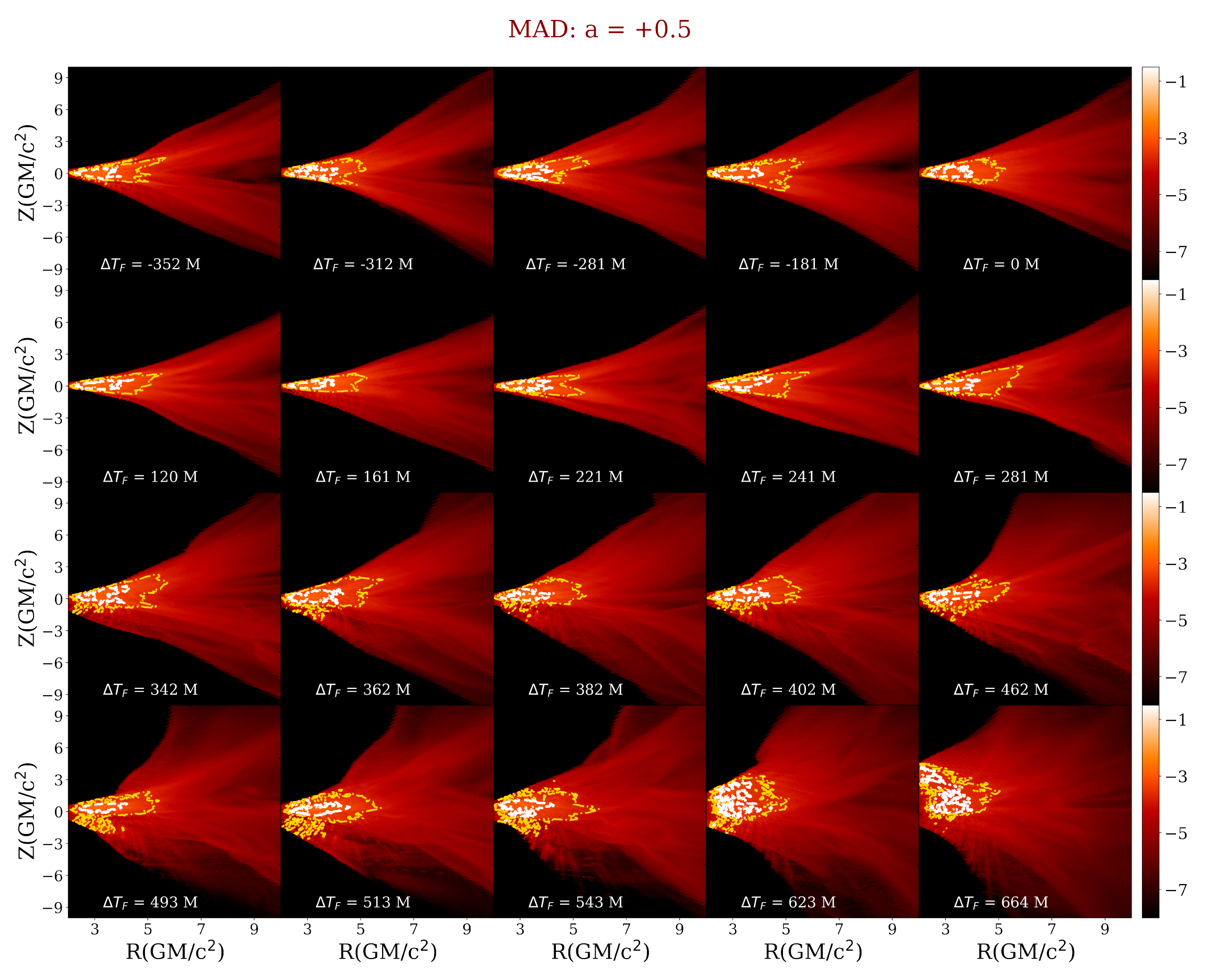}
\caption{Emission location in the vicinity of a flaring event in M87*.}
\label{Emission-location}
\end{figure*}
The flux maximum in MAD simulations shows an interesting spin dependence. While for the case with BH spin of $a = -0.94$, the flux shows a very oscillatory pattern, BH spins of $a = (-0.5, +0.5)$ show a more prominent looking flare. On the other hand, the BH spins of $a = (0.0, +0.94)$ do not seem to exhibit high-flux flares. Consequently, we argue that long-term monitoring of the M87* through the EHT will provide unique information about the BH spin. 

Since the flare looks very prominent in the case of $a = +0.5$, in the following we mainly focus on this case. It is also interesting to make a statistical analysis of individual flares for the case with BH spin of $a = -0.94$ and compare their features with each other. However, since they appear very close to each other, it is likely that the flaring effects somewhat get mixed from adjacent events, leading to a bias in the inferred polarimetric parameters near an actual flare. 
Consequently, we postpone this analysis to a future study and mostly limit our analysis to the case with $a = +0.5$. 

\subsection{Magnetic flux and accretion rate}
\label{B-field-accretion}
Having presented the intensity light curve for M87* here we examine if the intensity flare might be correlated with the mass accretion rate as well as the magnetic flux. We use the framework developed in \cite{Chatterjee:2022} to infer the mass accretion rate (their Eq. 3) and the magnetic flux (their Eq. 7). 
Figure \ref{light-curve-Magnetic-Flux-Accretion} presents the time evolution of the magnetic flux and the accretion rate for different GRMHD simulations. Overlaid in each panel we depict the time associated with the intensity flux with a dashed vertical orange line. It is explicitly seen that the magnetic flux is peaky near to the flare followed by a flux eruption. The accretion rate also gets to its local maxima at the flare time but it is less prominent. Finally, MAD simulations show a much clearer response to flares than the SANE models. 

\subsection{Damped Random Walk}
\label{DRW-fit}
Recent observations \citep{2023ApJ...951...93C} have suggested that the variability timescale of submillimeter emission provides a useful tool for studying the accretion to SMBHs. \cite{2009ApJ...698..895K}  has modeled the optical light curves of quasars as a damped random walk (DRW) stochastic process
finding that the characteristic variability timescale strongly correlates with the BH mass as well as the BH luminosity. This was then used as a standard tool for modeling the variability time-scale in SgrA* \citep{2022ApJ...930L..19W}, followed by a systematic modeling of low-luminosity 
AGNs with DRW \citep{2023ApJ...951...93C}. Motivated by this, in the following, we model our synthetic light curves with a DRW and estimate parameters including the characteristic timescale $\tau$, the mean flux $\mu$, and the variance of the DRW, $\sigma$. We present the final results here while leaving the details to Appendix \ref{DRW_Details}. 
Figure \ref{DRW-inferred-params} presents the 2$\sigma$ inferred DRW main parameters 
for different BH spins in both the MAD (blue) and SANE (purple) simulations. From left to right, we depict the characteristic timescale, mean flux, and the standard deviation for DRW, respectively. Overlaid in each panel, we depict the inferred parameters from SMA observations taken from \cite{2023ApJ...951...93C}. For the first time, we observe a mild spin dependency in $\tau$, extending the general expectation that $\tau$ is only a function of the BH mass and luminosity. Furthermore, $\tau$($\mu$) is above(below) the SMA observation highlighting the fact that SMA variability(flux) is a consequence of both the event horizon and the jet-based variabilities, while the GRMHD simulation results are only capturing the event horizon scale variabilities. 

In Figure \ref{DRW-MCMC} we present the posterior distribution corner plots
for a MAD simulation with a = + 0.5. Depicted for each parameter are the median and the (5, 95) percentiles. 

\subsection{Wavelet transformation}
\label{Wavelet-trans}
Next, we apply the wavelet transform to the light curve for the thermal model from 
Figure \ref{light-curve-M87}. This is a common approach in time-series analysis and allows us to make a detailed analysis of variations in the light curves by identifying the time and frequency components of the signal with relatively high resolution. This technique is very useful in detecting transients and flares in astrophysical systems. The time-frequency analysis of the flux intensity is presented in Figure \ref{Wavelet-Transformation} as a wavelet power spectrum. The plot displays the wavelet power spectrum as colors in the frequency-time plane, with brighter colors indicating higher power. This figure highlights the temporal evolution of frequency components and their relative strengths during the observed interval.
Flux flares appear as regions with higher power. Interestingly, the flux flare for the case with $a = +0.5$ is associated with the highest wavelet power spectrum, while the occurrence of multiple flaring events in $a = -0.93$ makes it  
more challenging to separate them in the power spectrum as well. 

\begin{figure*}
\center
\includegraphics[width=0.99\textwidth]
{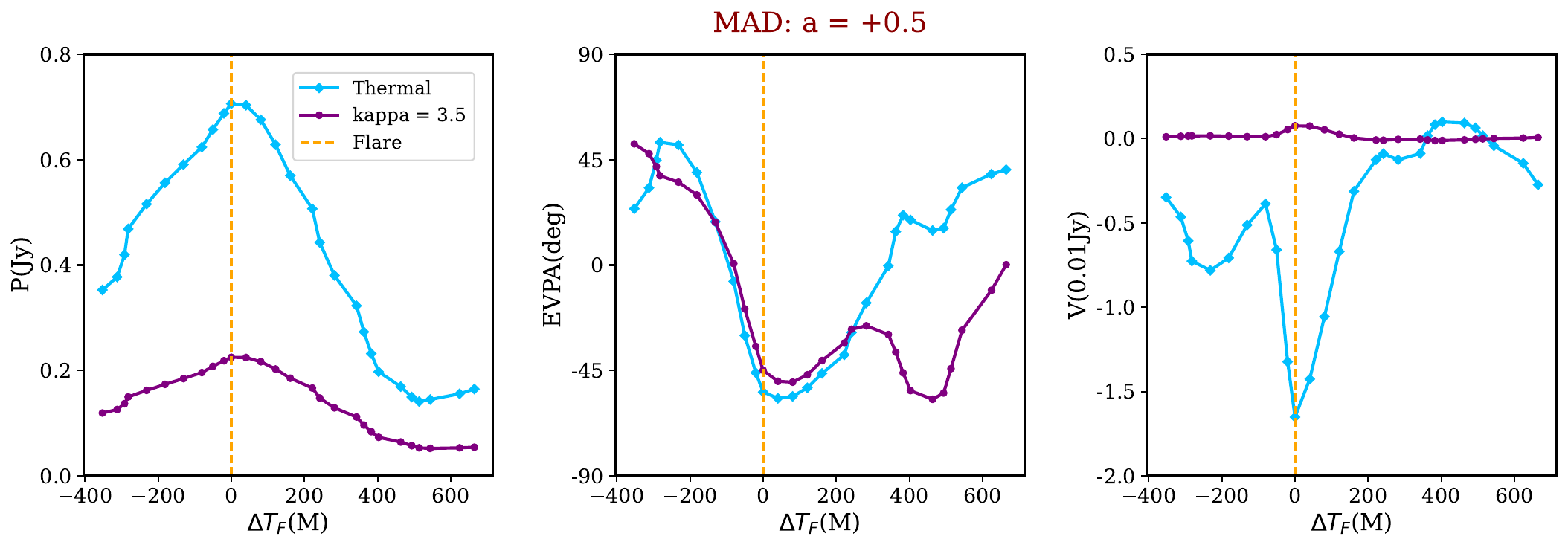}
\caption{The time evolution of the average linear polarization(P), the EVPA, and 100 times the circular polarization is depicted from left to right. In each panel, the flaring moment is presented as a dashed black line. Overlaid in each plot we present both the thermal and non-thermal $\kappa$ model with the orange and blue lines, respectively. Both the linear and the circular polarization show sharp variations in the vicinity of the flux flare. The EVPA changes its sign close to the flare as well.  }
\label{Stokes_Flare}
\end{figure*}

\subsection{BH image in the vicinity of a flare} \label{flare-image}
Having focused on flares for the case with a BH spin of $a = +0.5$, here we present the BH image in the vicinity of an actual flaring event. In Figure \ref{Image-Flare-ap5} we show the images for M87* close to a flare. It is inferred from the plot that the intensity starts with a gradual increase before the flare and that is followed by a flux eruption \citep{2022Galax..10..107G} after that where the flux expands from the inner side to the outer part of the image, leading to a significant flux diminish of about 90\% of the corresponded flux at the flare location. The flare corresponds to the intensity peak at $I = 1.57$ Jy at $\Delta T_F$ = 0 M. Time is shifted with respect to the flare's time meaning that negative times refer to moments before the flare while positive times correspond to the time after the flare. 

Overlaid in each panel, we also present the electric vector polarization angle, hereafter EVPA, ticks. It is inferred that the EVPA ticks change their orientation along the fluxing event. In Section \ref{Polarimetric-Flare-Image} we will quantify these variations.

\subsection{Tracing the emission location in the vicinity of flare} \label{Emission-location-BH}
Since the BH image is significantly different around a flaring event, it is intriguing to infer the source of the emission. In \cite{2023ApJ...950...38E} we computed the emission location for the time-averaged images in a wide range of simulations. Here we extend that analysis and infer the azimuthally averaged emission location in the vicinity of a flare in M87*. Figure \ref{Emission-location} depicts the emission location for 20 snapshots near a flaring event for the case with $a = +0.5$. To facilitate the comparison with the BH image from Figure \ref{Image-Flare-ap5} we use the same timing. In each panel, we highlight the emission location for the top 20\% and 50\% emission with the white(dashed) and yellow(dotted-dashed) lines, respectively. It is inferred that while during the flaring event, the majority of the emission is focused on the equatorial plane, during the flux eruption phase the dominant emission is expanded beyond the equatorial plane and every cell in the $R-Z$ plane somehow contributes to the synchrotron emission. It is also seen with the in-out expansion of the white and yellow lines from close to $Z = 0$ to beyond the equatorial plane.

\section{Polarimetric analysis in the vicinity of a flare: Image space} 
\label{Polarimetric-Flare-Image}
Since the EVPAs demonstrate some levels of variation in the vicinity of an actual flare, here we conduct an in-depth polarimetric analysis near a flux flare. 

Figure \ref{Stokes_Flare} depicts the polarimetry in the vicinity of a flare. From left to right we present the linear polarization, the EVPA, and the circular polarization. In each panel, the orange and blue lines present the thermal and non-thermal $\kappa =3.5$ model, respectively. The flare moment is depicted as a dashed black line. We uncover evidence of higher linear and circular polarization during the flaring event. Since the polarization is higher in the thermal model compared to the $\kappa$ model, the changes are more prominent in the thermal model. The EVPA patterns in the middle panel show an interesting pattern where the EVPA changes its sign in the vicinity of the flare and the sign change looks the same for both of the thermal and $\kappa$ models.

The sign of the circular polarization differs between the thermal and $\kappa$ models. While in the thermal model  circular polarization is mainly negative, it rises above zero for the $\kappa$ model, yet its amplitude is much lower in the $\kappa$ model compared to the thermal case. 
\begin{figure*}
\center
\includegraphics[width=0.9\textwidth]{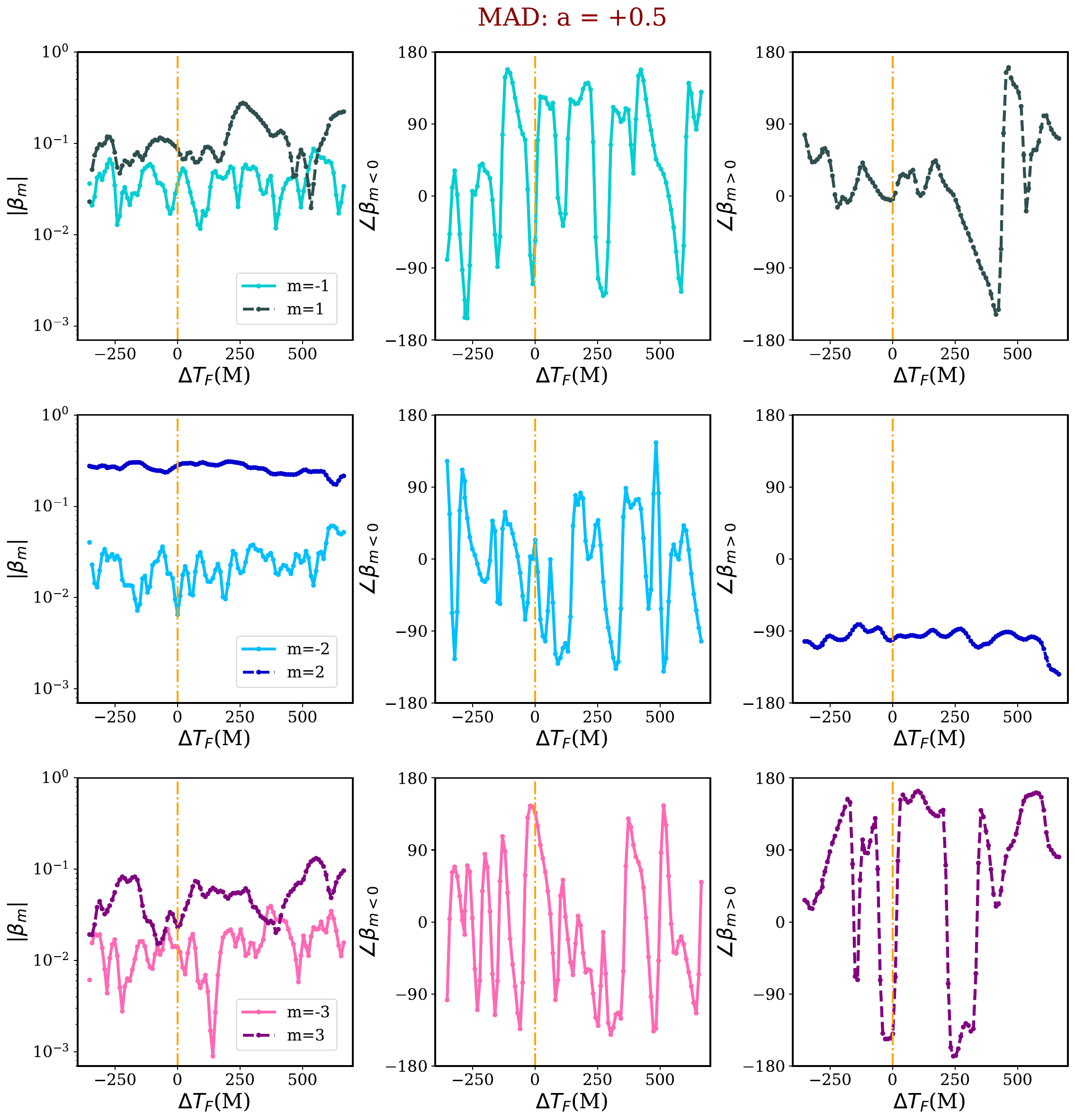}
\caption{The time evolution of six $\beta_m$ modes near to a flux flare for the case of $a = +0.5$. While $m=2$ does not show any time variations in the vicinity of the actual flux flare, other modes demonstrate more prominent variations during the flux flare.  }
\label{Betam_flare}
\end{figure*}

Since the average EVPA shows a sign shift in the vicinity of an actual flare, here we 
utilize using the azimuthal decomposition of  the polarized BH image in terms of complex $\beta_m$ modes as: 
\begin{align}
    \beta_m &=\dfrac{1}{I_{\rm tot}} \int\limits_{0}^{\infty} \int\limits_0^{2 \pi} P(\rho, \varphi) \, e^{- i m \varphi} \; \rho \mathop{\ud\varphi}  \mathop{\ud\rho},\\
    I_{\rm tot} &= \int\limits_{0}^{\infty} \int\limits_0^{2 \pi} I(\rho, \varphi) \; \rho \mathop{\ud\varphi} \mathop{\ud\rho}.
\end{align}
Figure \ref{Betam_flare} depicts the time evolution of six $\beta_m$ modes. From the top to bottom rows we present $m = \pm 1$, $m = \pm 2$, and, $m = \pm 3$, respectively. In each row from the left to right we  
depict the amplitude of $\beta_m$, the phase of $\beta_m$ for negative modes and the positive modes, respectively. It is inferred that both the amplitude and the phase of the $\beta_2$ do not show any significant variations in the vicinity of the flaring event.  Other $m$ modes, on the other hand, demonstrate some levels of time evolution. For example, the phase of $m = -2$ shows a diminishing profile near the flare. On the other hand, the phases of $m=-1$, $m=-3$, and, $m=3$ demonstrate sudden variations in the vicinity of the flare.  

\begin{figure*}
\center
\includegraphics[width=0.9\textwidth]
{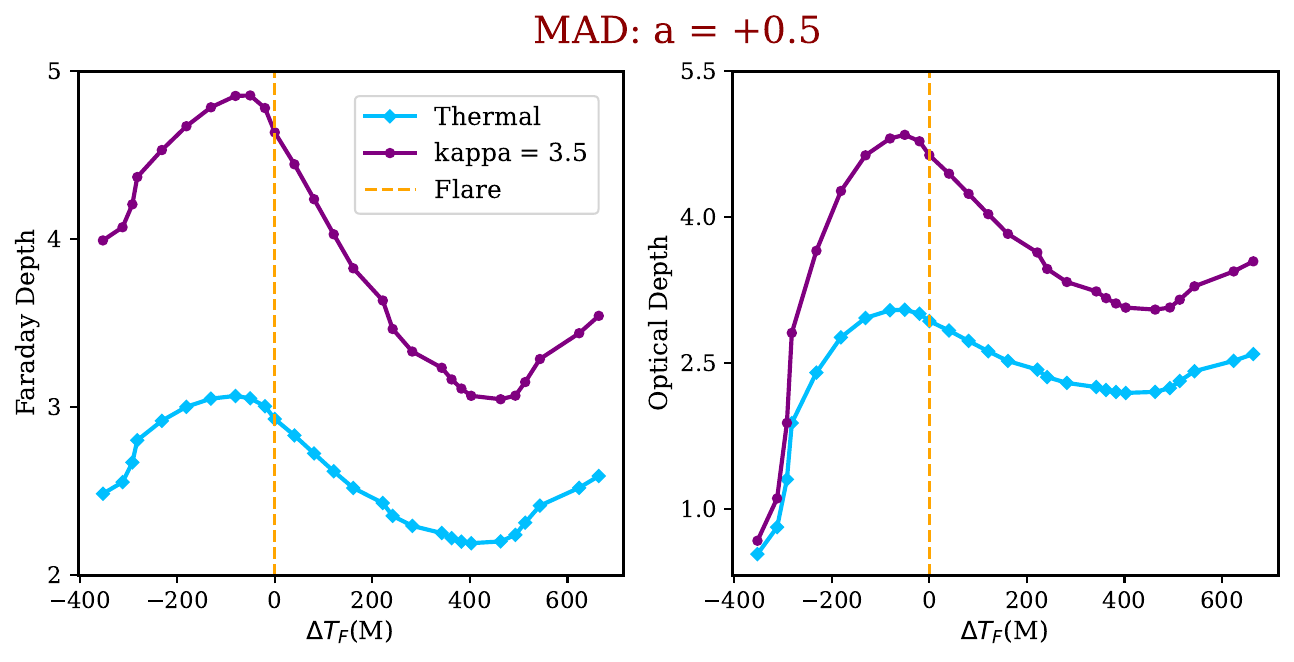}
\caption{The time evolution of the Faraday Depth (left panel) and the Optical Depth (right panel) in the vicinity of a flaring event for the case with BH spin of a = +0.5. From the plot, it can be seen that both reach a peak in the vicinity of the BH flaring event. }
\label{Faraday_Optical_Flare}
\end{figure*}

Figure \ref{Faraday_Optical_Flare} presents the time evolution of the Faraday Depth (left panel) and the Optical Depth (right panel) in the vicinity of a flaring event for the case with BH spin of a = +0.5. From the plot, it can be seen that both reach a peak in the vicinity of the BH flaring event.

\section{Polarimetric analysis in the vicinity of a flare:  Visibility space }
\label{Polarimetric-Flare-Image-Visibility-space}
In this section, we extend the analysis above in the image space and probe the signatures of a flaring event in the visibility space. In Sec. \ref{Visibility-amplitude-flare} we study the visibility amplitude signatures, while in Sec. \ref{EB-correlation-flare} we quantify the signatures of a flaring event through the analysis of the phase of the $EB$-correlation function. 

\begin{figure*}
\center
\includegraphics[width=0.99\textwidth]{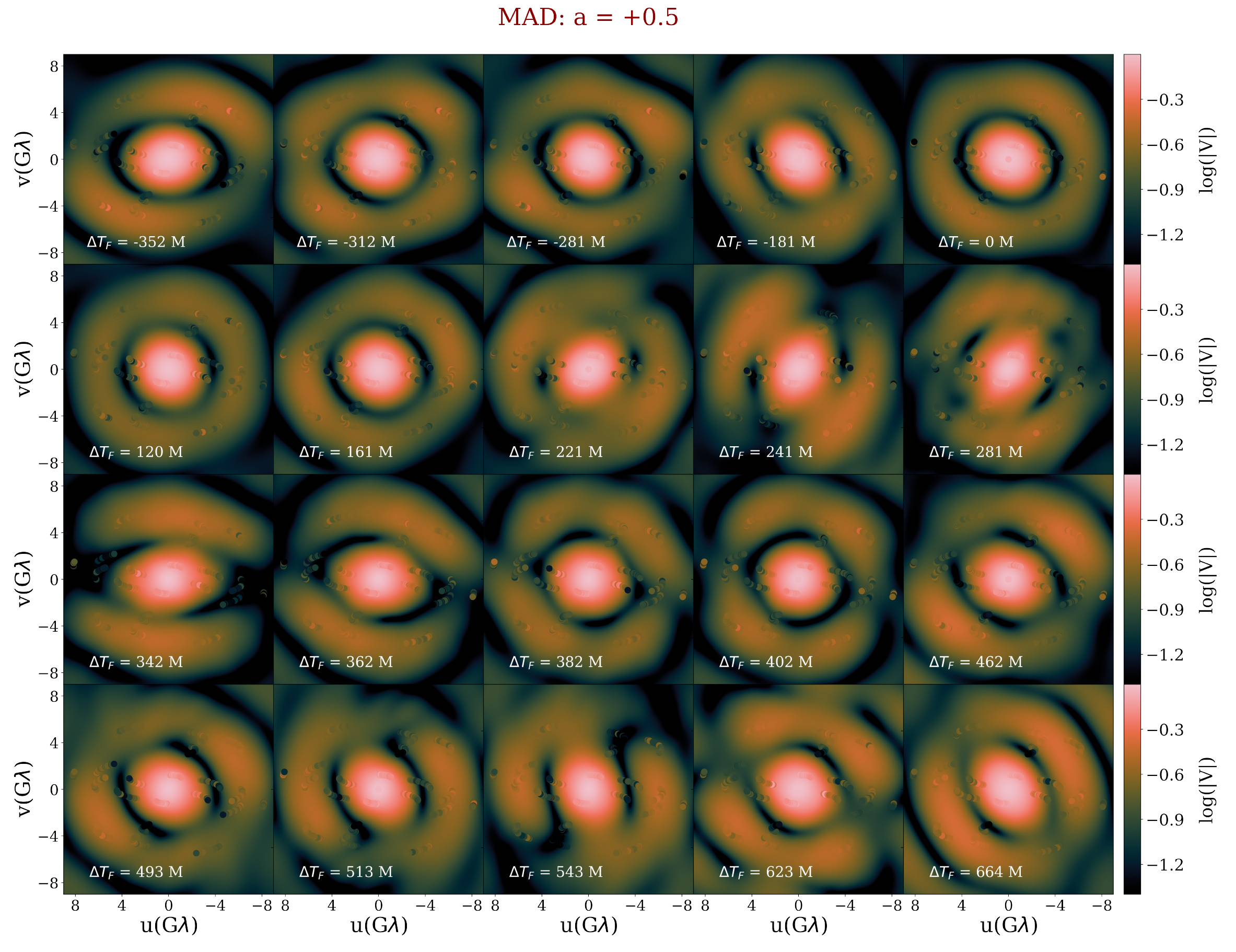}
\caption{The logarithm of the visibility amplitude in the vicinity of a flaring event in M87*. }
\label{Image_Flare_vis}
\end{figure*}
\subsection{Visibility amplitude close to a flare} \label{Visibility-amplitude-flare}
In EHT observations \citep{PaperIII}, the visibility function captures the mutual coherence of the electric field between the two ends of the baseline vector, projected onto the plane of the sky. In an ideal interferometer, this function directly corresponds to a sampled Fourier component of the sky's brightness distribution: 
\begin{equation}
\label{real-Fourier}
V(u, \theta) = \int\int I(\rho, \phi) e^{-2i\pi \rho u \cos{(\theta - \phi)}} \rho d\rho d\phi.
\end{equation}
where $I$ represents the intensity in the image space, while $u$
denotes the spatial frequency of the Fourier component, defined by the projected baseline and measured in units of the observing wavelength.
The visibility function is an excellent way to capture image asymmetries in the visibility space. Motivated by this, here we compute the map of the visibility amplitude in the vicinity of a flaring event. 

Figure \ref{Image_Flare_vis} presents the logarithm of the visibility amplitude in visibility space $(u,v)$ near to the flaring event. While the map looks very symmetric at the flare, its morphology significantly varies and it experiences some levels of re-orientation and twistiness in the vicinity of an actual flare. Consequently, we conclude that the visibility amplitude can be used as a diagnostic of a flare. 

\begin{figure*}
\center
\includegraphics[width=0.99\textwidth]{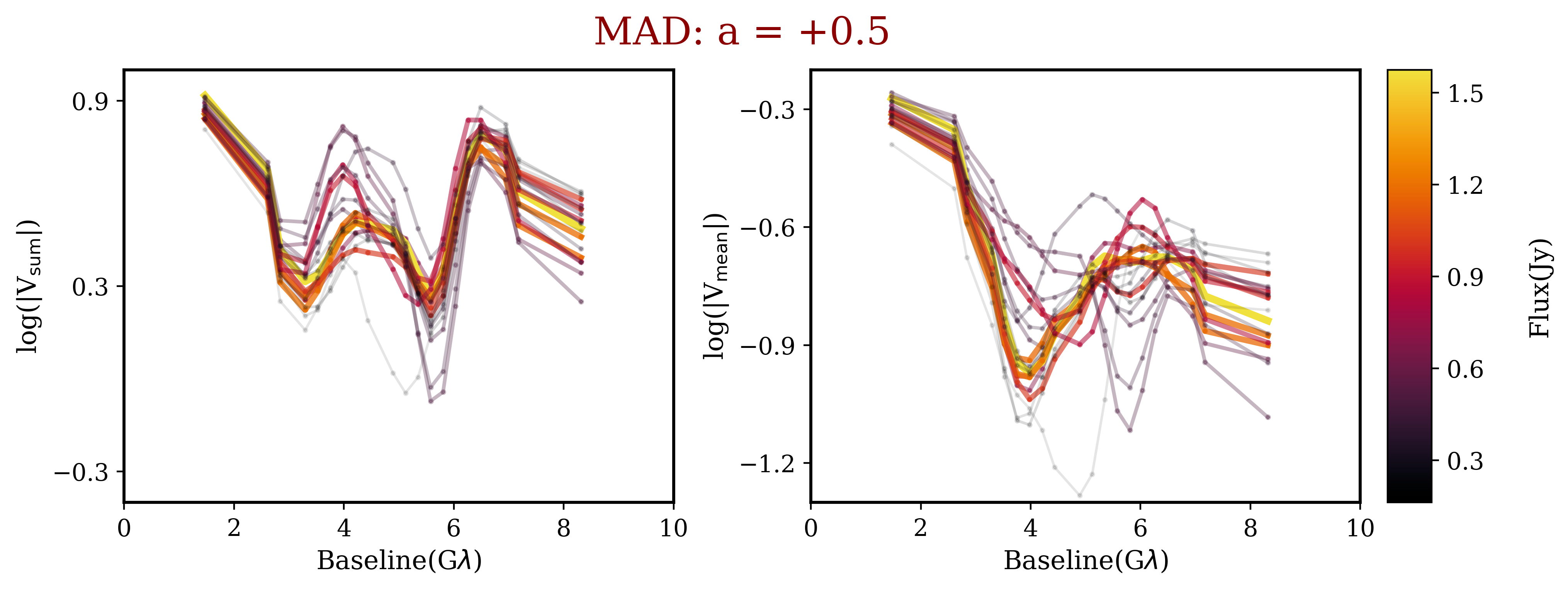}
\caption{The logarithm of the visibility amplitude vs. the baseline lengths in the vicinity of a flaring event, corresponds to the maximum flux. }
\label{Flare_Vis_Baseline}
\end{figure*}
To identify which baselines are most affected by the flare, Figure \ref{Flare_Vis_Baseline} presents a one-dimensional representation of the logarithm of the visibility amplitude as a function of baseline length, defined as 
Baseline $=\sqrt{(u^2 + v^2)}$. To evaluate the visibility amplitude, we divide the visibility space into grids and compute the summation (left panel) and mean value (right panel) of the visibility amplitude within each grid. From the plot, it is inferred that while the short baselines exhibit no significant signatures, both the intermediate and long baselines display pronounced variations in visibility amplitude close to the flaring event. However, the precise values of these variations depend on the specific metric used for the analysis.
\begin{figure*}[t!]
\center
\includegraphics[width=0.93\textwidth]{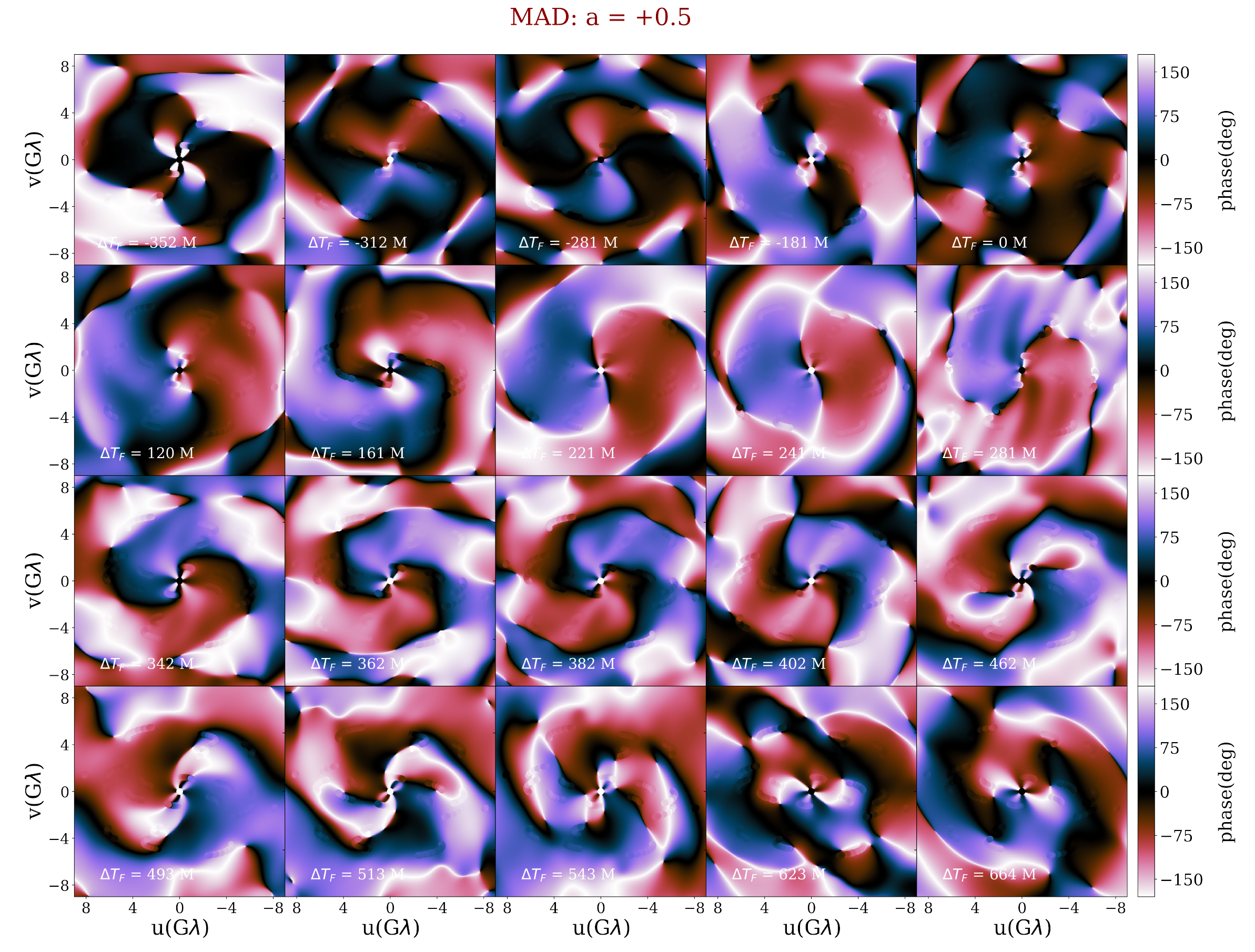}
\caption{The $EB$-correlation phase map in the visibility space exhibits notable variations near a flaring event. The correlation phase undergoes significant changes during different stages of flux enhancement and subsequent flux eruption, offering a reliable method to trace the transient phase of a flaring event.}
\label{Image_Flare_EB_phase}
\center
\includegraphics[width=0.91\textwidth]{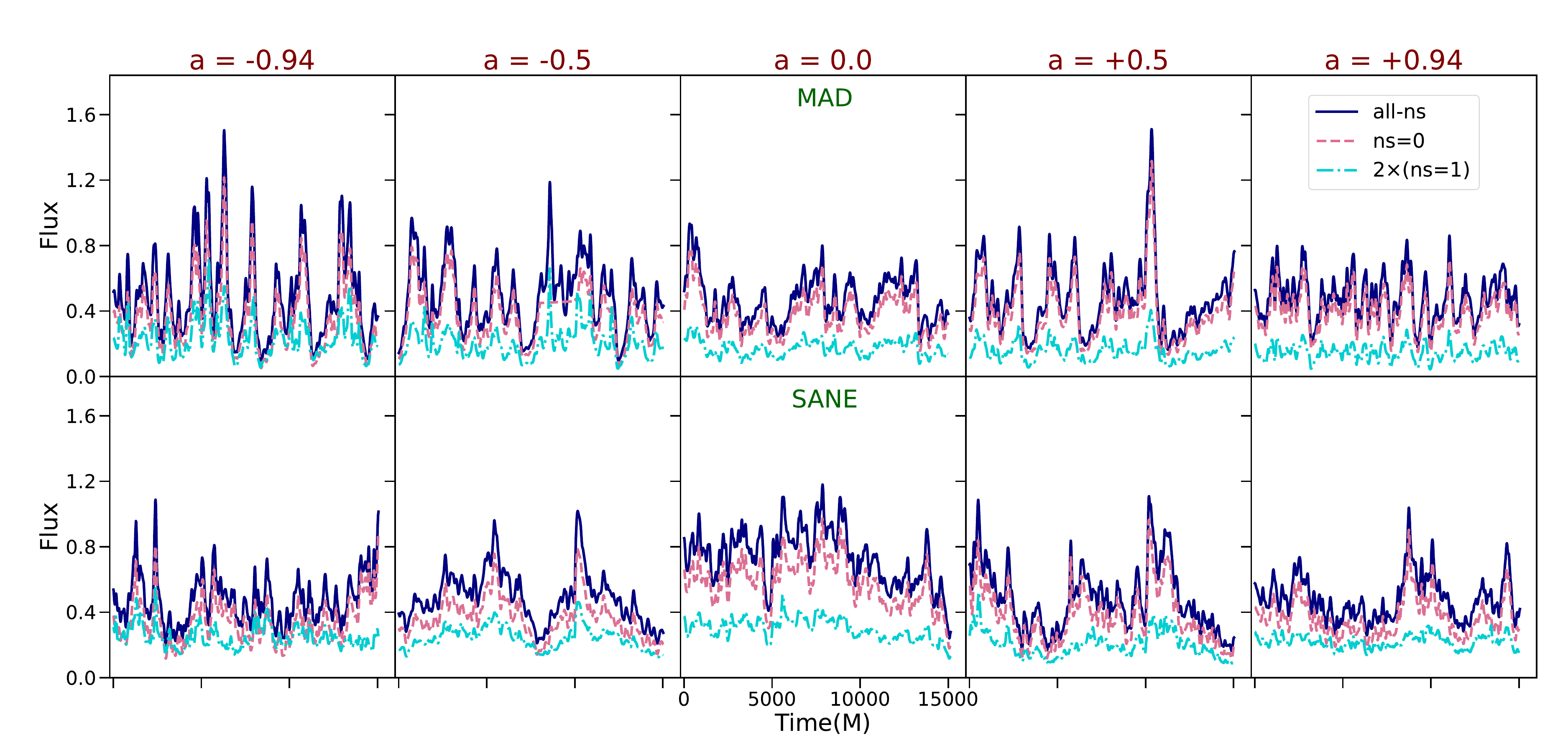}
\caption{Light curve for M87* in MAD and SANE simulations. Blue, pink, and cyan lines refer to the full emission, n=0 sub-ring, and n=1 sub-rings, respectively. For better scaling, the flux values for $n=1$ sub-ring are multiplied by a factor of 2. }
\label{light-curve-subring-fig}
\end{figure*}

\begin{figure*}[th!]
\center
\includegraphics[width=0.97\textwidth]
{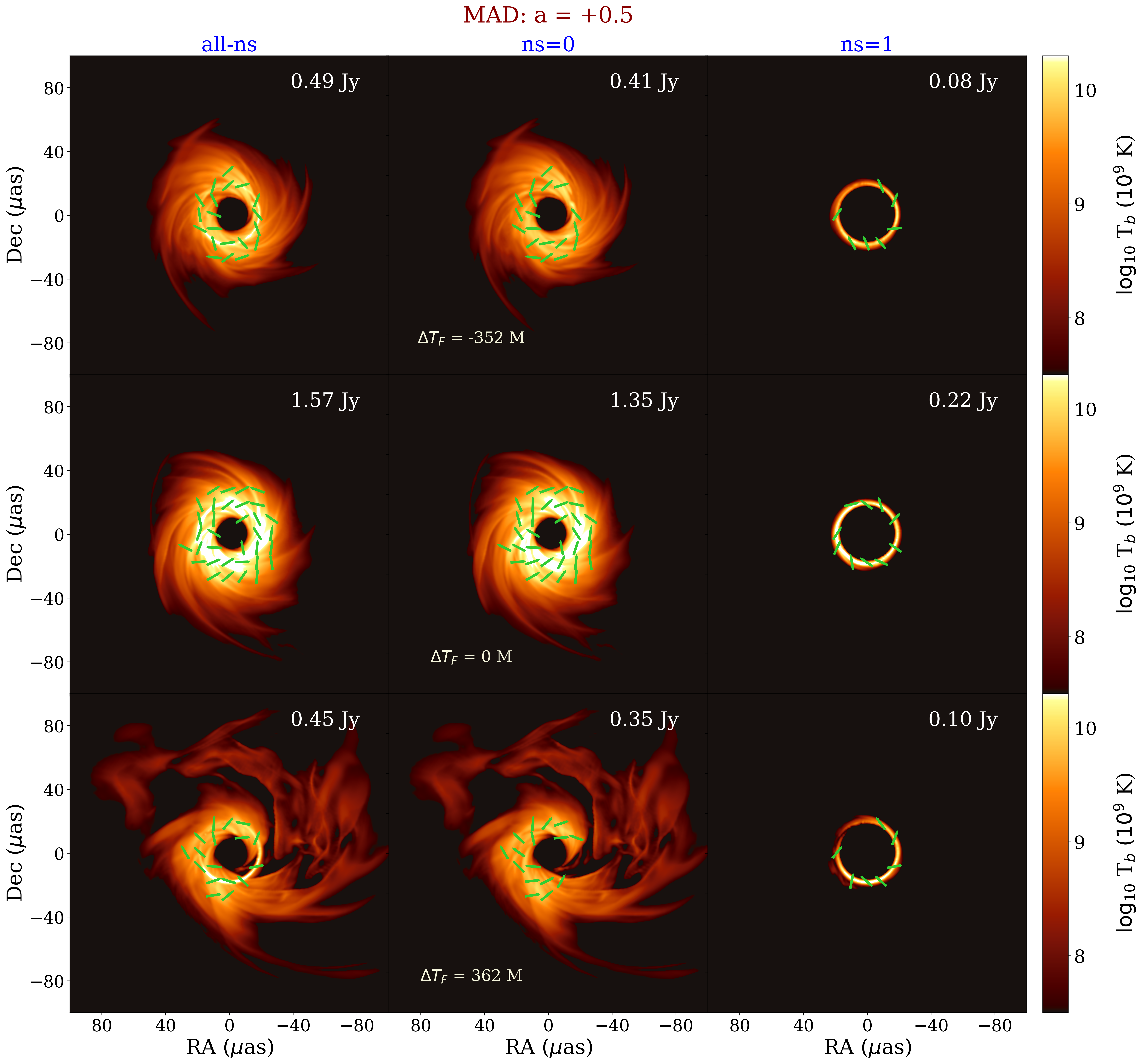}
\caption{M87* image for three snapshots in the vicinity of a flaring event for a=+0.5. In each row from the left to right we study the cases with all-ns, ns=0, and ns=1, respectively. }
\label{BH-Image-Subring}
\end{figure*}
\subsection{EB correlation near to a flare} \label{EB-correlation-flare}
Next, we study the signatures of this flaring event in the phase of the $EB$-correlation in the visibility space. In \cite{2023ApJ...955....6E} we developed a novel technique to infer the $EB$-correlation function in visibility space as: 
\begin{equation}
\label{EB-correlation}
\rho_{\mathcal{EB}}(u,v) \equiv \frac{E(u, v) B^{*}(u, v)}{\sqrt{E(u,v)  E^*(u,v)} \sqrt{B(u,v) B^*(u,v)}},
\end{equation}
The phase of this correlation function is defined as: 
\begin{equation}
\label{EB-phase}
\theta(u,v) \equiv \arctan{\left( \frac{Im(\rho_{\mathcal{EB}}(u,v))}{Re(\rho_{\mathcal{EB}}(u,v))} \right)}.
\end{equation} 
\begin{figure*}
\center
\includegraphics[width=1.01\textwidth]
{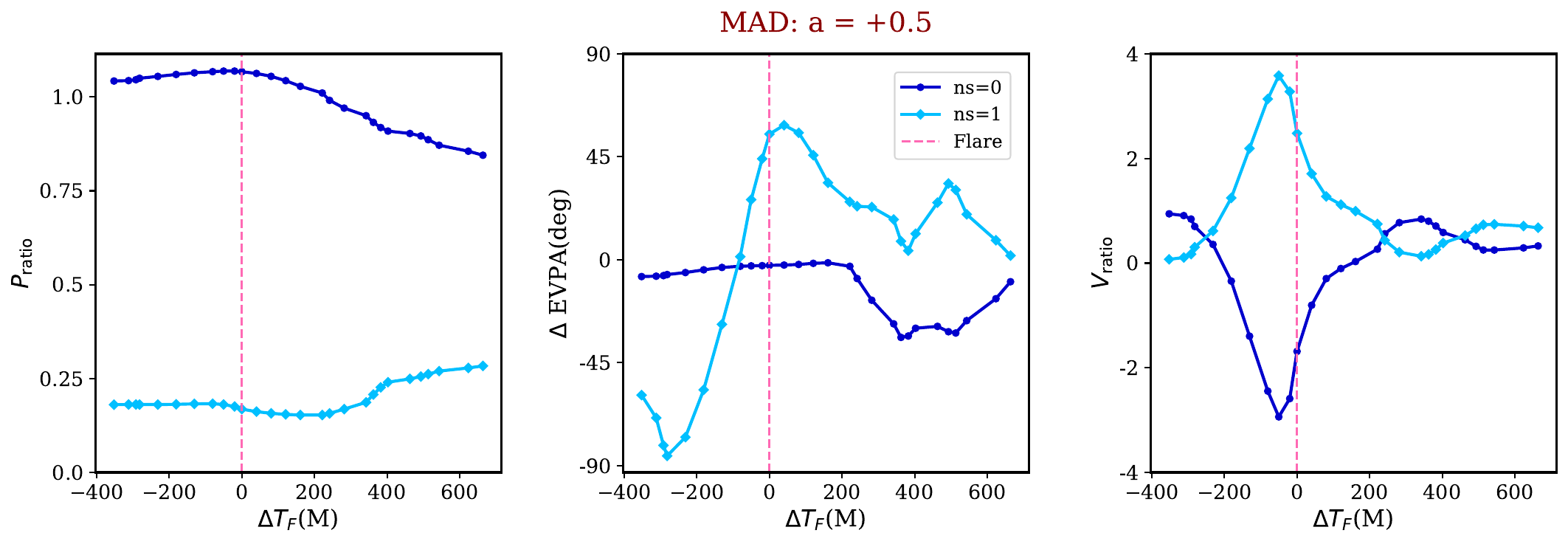}
\caption{The ratio of the linear polarization of different sub-rings to the full image (left panel), the difference between the EVPAs of different sub-rings compared to the full image (middle panel), and the ratio of the circular polarization between different sub-rings and the full image (right panel) in the vicinity of a flaring event.  }
\label{light-curve-all}
\end{figure*}
\begin{figure*}
\center
\includegraphics[width=1.01\textwidth]
{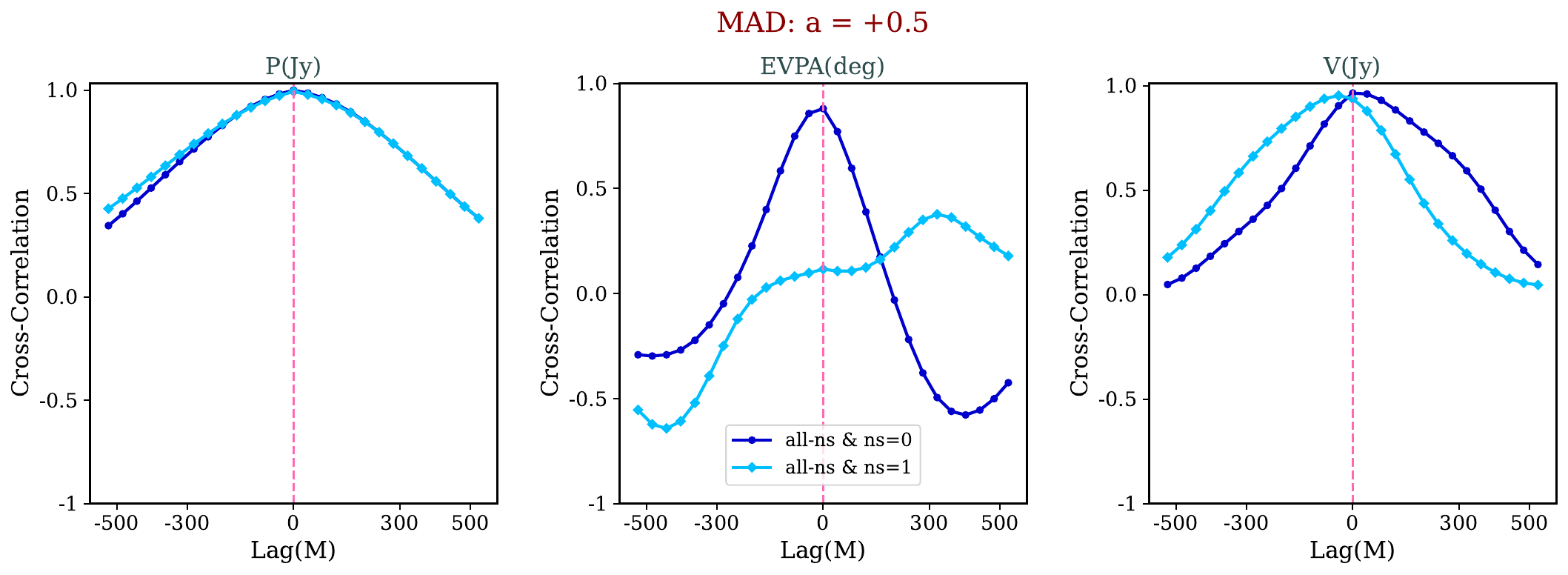}
\caption{The cross-correlation between the all-ns vs. ns=0 (blue) and ns=1 (cyan) vs. The time lag for the linear polarization (left panel), the EVPA (middle panel) and the circular polarization (right panel, respectively.)}
\label{cross-correlation}
\end{figure*}
Here we use this approach and compute the correlation phase in the vicinity of a flaring event in M87*. 

Figure \ref{Image_Flare_EB_phase} depicts the $EB$-correlation phase map close to the flaring event. From the plot it is evident that the correlation phase significantly varies between the phase of flux enhancement (first row) followed by the gradual flux diminishing (second row), and the flux eruption 
(third-fourth rows). Comparing individual panels from this figure with their corresponding values from Figure \ref{Image-Flare-ap5} it is evident that for all morphological variations in the BH image, their corresponding $EB$-correlation phase also alters. Consequently, we conclude that the $EB$-correlation phase provides a robust approach to trace the transient phase in a flaring event.

\section{Tracing the flare in photon ring}
\label{sub-ring}
In the framework of general relativity, BH images contain a feature known as the "photon ring" \citep{2020SciA....6.1310J}, consisting of an infinite sequence of sub-rings. These sub-rings are characterized by the number of photon orbits completed around the BH. Each sub-ring contains novel information about the BH 
astrophysics and the BH spin. In the following sections, we make an in-depth study of the first few  sub-rings including, ns = 0 and ns =1 and compare them against the full image, hereafter referred to as all-ns.

\subsection{Light curve sub-rings}
\label{Light-curve-subring}
We start by looking at the intensity light curve from different sub-rings. Figure \ref{light-curve-subring-fig} presents the light curve for MAD (top row) and SANE (bottom row) simulations. In each row from left to right we increase the BH spin. In each panel, we compare the flux intensity for three cases, including all-ns, ns=0 and ns=1. As the photon ring index increases, the photon ring becomes thinner and thus the flux diminishes. To facilitate the comparison we multiplied the flux for ns=1 by a factor of two. 

From the plot, it is evident that different sub-rings mainly follow the time variability structures of the flux intensity, although the structures in MAD simulations are more similar than in SANE simulations. 

Given this general picture of the variability structure of different sub-rings, hereafter we restrict our study to the case of MAD simulations with the BH spin of a = +0.5. 

\subsection{Polarimetric signatures of flares in photon sub-rings: Image space}
\label{Image-subring-pol}
Next, we study the polarimetric signatures of sub-rings near to the flare. Figure \ref{BH-Image-Subring} presents the BH images for three snapshots, taken from Figure \ref{Image-Flare-ap5}. In each row, the left-to-right columns refer to the all-ns, ns=0, and ns=1 cases, respectively. The image morphology and the EVPA tick patterns are very similar between the all-ns and ns=0, while they are substantially different for the case with ns=1. In more detail, the EVPA ticks are rotated between all-ns and ns=1 cases. 

\begin{figure*}
\center
\includegraphics[width=1.01\textwidth]
{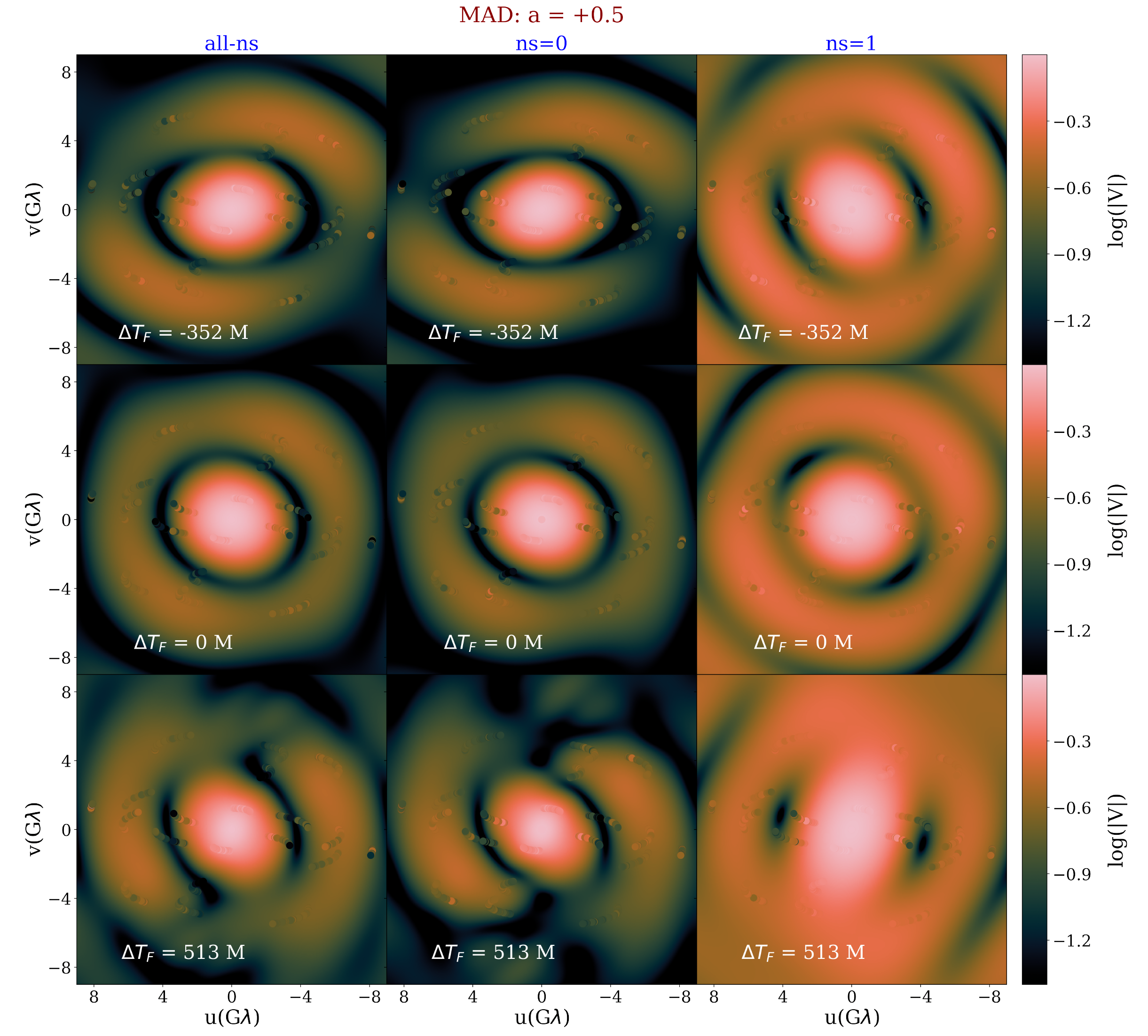}
\caption{The map of the logarithm of the visibility amplitude for three snapshots in the vicinity of a flaring event for different photon sub-rings. In each row from the left to right we present the cases with all-ns, ns=0, and ns=1, respectively. While the morphology of all-ns and ns=0 are fairly similar, ns=1 presents different fingerprints.}
\label{Visibility-subring}
\end{figure*}
The comparison between all-ns polarimetric parameters and ns=0 and ns=1 is 
further quantified in Figure \ref{light-curve-all}. Here, from left to right we present the ratio of linear polarization, the difference between the EVPA ticks, and the ratio of the circular polarization between the all-ns with ns=0 (depicted by blue line) and all-ns compared to ns=1 (shown as cyan line), respectively. In all of these cases, there is a transition in the polarized parameters before and after the flare. While there is a delay in the transition for the linear polarization ratio, the circular polarization presents a sharp variation during the flare. Interestingly, while in the linear polarization, all-ns is dominant over both ns=0 and substantially for ns=1, the opposite is true for the circular polarization case. Near to the flare, the circular polarization for ns=0 and ns=1 greatly exceeds the case of all-ns. However, since their sign is opposite their contribution almost cancels out leaving us with smaller circular polarization for the full image. This implies that the futuristic VLBI will be important for detecting the circular polarization near to the flare. 

The EVPA difference between all-ns cases with ns=1 presents a twisted pattern of above 90 degrees from before to after the flaring event. 

\begin{figure*}
\center
\includegraphics[width=1.01\textwidth]
{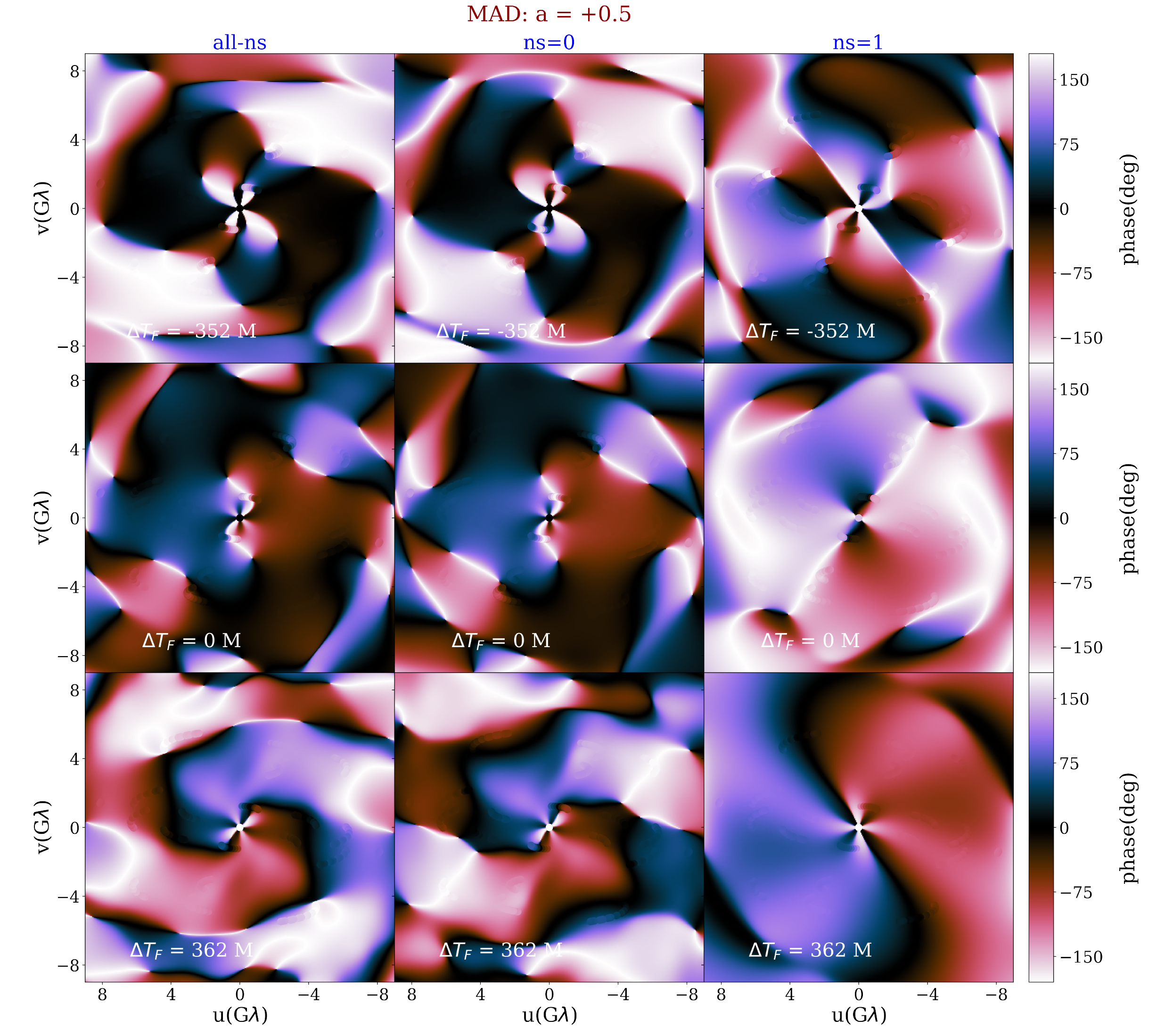}
\caption{The map of the $EB$-correlation phase for three snapshots in the vicinity of a flaring event for different photon sub-rings. In each row from the left to right we present the cases with all-ns, ns=0, and ns=1, respectively. While the morphology of the phase map between all-ns and ns=0 are fairly similar, ns=1 presents different fingerprints.}
\label{EB-correlation-subring}
\end{figure*}
Next, we compute the  normalized cross-correlation function as:
\begin{equation}
Cor(\mathrm{Lag}) = \sum_{t} \frac{A(t+\mathrm{Lag})B(t)}{|A(t)|~|B(t)| }
\end{equation}
where $A$ refers to the polarimetric data from all-ns case, while $B$ describes either ns=0 or ns=1, respectively. Additionally, $\mathrm{Lag}$ refers to the time lag in the unit of $M$. 
Figure \ref{cross-correlation} depicts the cross-correlation function between the all-ns vs. ns=0 (blue) and ns=1 (cyan) vs. the time lag for the linear polarization (left panel), the EVPA (middle panel) and the circular polarization (right panel, respectively.)
While the cross-correlation for the linear and the circular polarization has a peak at zero time lag, its peak is delayed for the EVPAs associated with ns=1. This may mean that the EVPA ticks make some rotations for ns=1, due to the time-delay effect, and that causes a shift in their maximum correlation which does not exist for the amplitude of the linear polarization and the circular polarization. Furthermore, depending on the time lag value, the EVPAs might be anti-correlated, which does not exist for linear and circular polarizations. 

\subsection{Polarimetric signatures of flares in photon sub-rings: Visibility space}
\label{Visibility-subring-pol}
Here we study the polarimetric signatures of a flaring event from the sub-rings in visibility space. Figure \ref{Visibility-subring} presents the logarithm of the visibility amplitude in the (u,v) space for three snapshots in the vicinity of a flaring event. In each row from left to right, we present the case with all-ns, ns=0, and ns=1. It is seen that while the distribution of the visibility amplitude for all-ns and ns=0 are fairly closed, ns=1 has completely different fingerprints. Furthermore, its elongation is also not the same as for the case with all-ns and ns=0. 
While ns=1 carries a portion of the flux, since this is thin, its higher flux is expanded further in visibility space. This makes it easier to detect variabilities for ns=1 in visibility space.  
\begin{figure*}
\center
\includegraphics[width=0.99\textwidth]{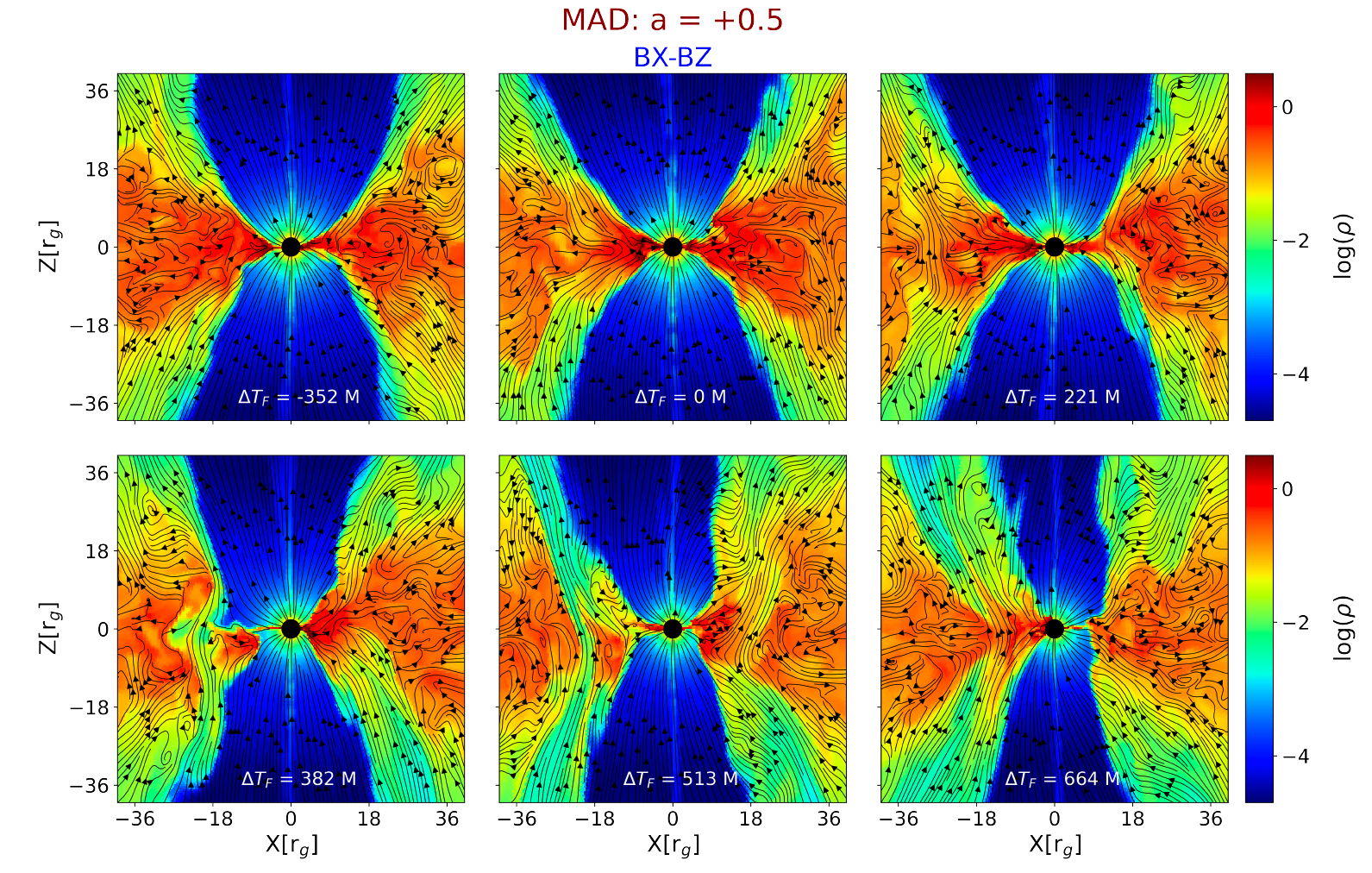}
\caption{The projected magnetic field in the X-Z plane in the vicinity of the flaring event.} \label{B-fieldXZ}
\includegraphics[width=0.99\textwidth]{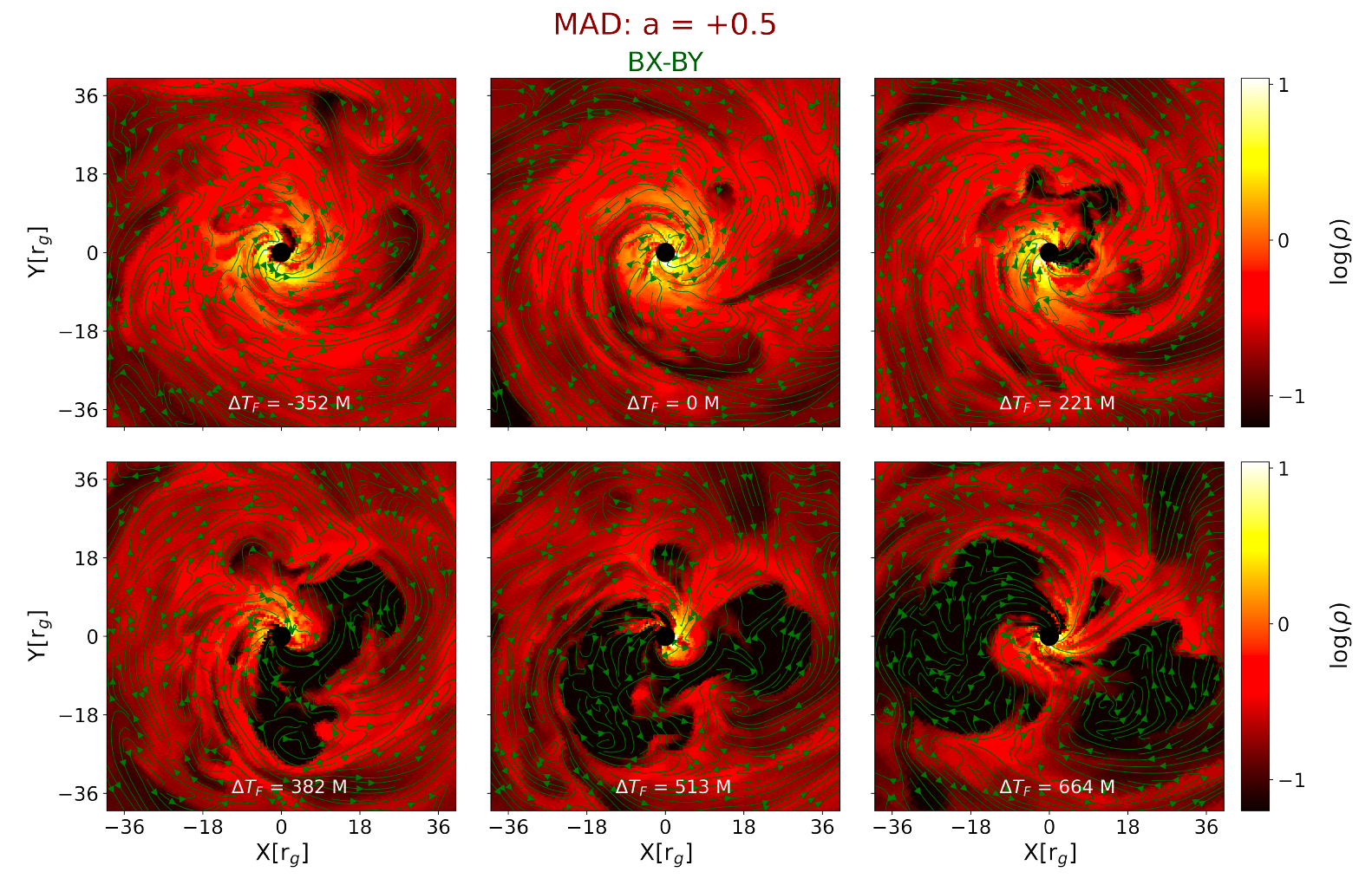}
\caption{The projected magnetic field in the X-Y plane in the vicinity of the flaring event.}
\label{B-fieldXY}
\end{figure*}
 
Finally, we study the imprints of different photon sub-rings in the phase of the $EB$-correlation function. Figure \ref{EB-correlation-subring} presents the $EB$-correlation phase map in visibility space for three snapshots in the vicinity of a flaring event for a=+0.5. As expected while the correlation phase maps remain quite similar between all-ns and ns=0 cases, they present completely distinct features for the case with ns=1. That includes some flips in the phase sign as well as substantial differences in the morphology of the phase map from ns=1. 

\begin{figure*}
\center
\includegraphics[width=0.99\textwidth]{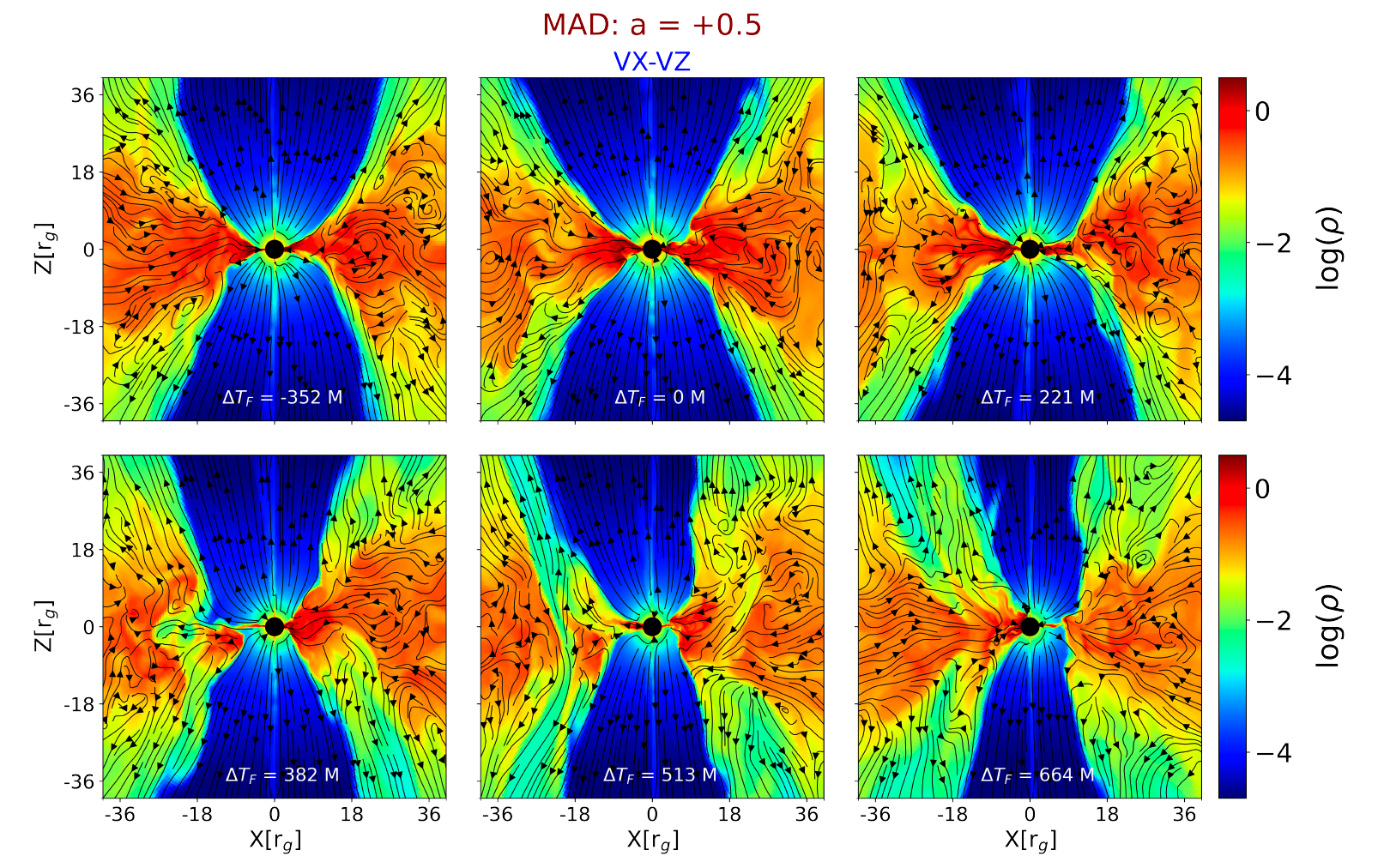}
\caption{The projected velocity field in the X-Z plane in the vicinity of a flaring event.}
\label{V-fieldXZ}
\includegraphics[width=0.99\textwidth]{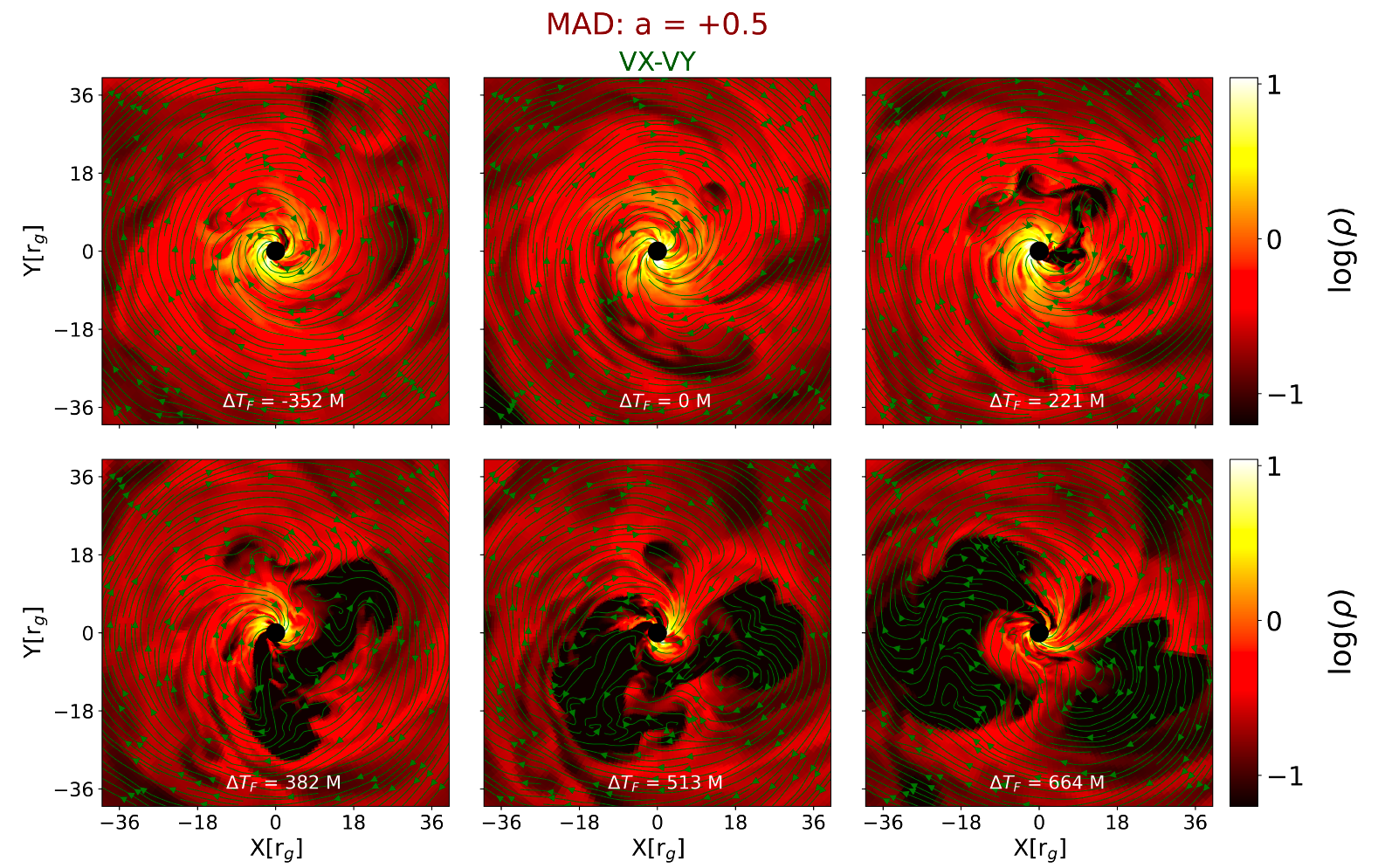}
\caption{The projected velocity field in the X-Y plane in the vicinity of a flaring event.}
\label{V-fieldXY}
\end{figure*}
\section{Magnetic field morphology close to a flare}
\label{Magnetic-field-flare}
Having studied different observational signatures of a radio flaring event in M87*, here we extend this analysis and probe the variations in the magnetic field morphologies in the vicinity of a flaring event. Figure \ref{B-fieldXZ} and Figure \ref{B-fieldXY} present the 2D projected magnetic field in the X-Z and X-Y planes, respectively. We take a sub-sample of snapshots from Figure \ref{Image-Flare-ap5} and showcase the structure of the magnetic field in these snapshots. In the first row, the first column presents the structure before the flare, the second panel at the time of the flare and the third panel right after the flare and before the beginning of the subsequent flux eruption period. This is followed by the second row where we depict the magnetic field structure early on in the flux eruption (left panel), toward the maximum stage of the flux eruption (middle panel), and after the end of the flux eruption (right panel) where the flux is being replenished by further accretion to the BH. 

It is inferred from the plot that while there are some magnetic loops and magnetic reconnection just before the flare, the structure of the magnetic field becomes simpler at the moment of the followed by more reconnection after the flare. Finally, at the flux eruption stage the short-term magnetic loops are removed,
resulting in a more coherent magnetic field structure both in the X-Z and X-Y projections.

\section{Velocity field morphology close to a flare}
\label{Velocity-field-flare}
Here we study the velocity field structure in the vicinity of a flaring event. 
Figure \ref{V-fieldXZ} and \ref{V-fieldXY} present the projected velocity field in the X-Z and X-Y planes, respectively. 
From Figure \ref{V-fieldXZ} it is inferred that along the Z direction and close to the center there are inflowing and outflowing currents that are separated from each other. Furthermore, Figure \ref{V-fieldXY} shows that while the flow has a clockwise motion in the X-Y plane near the flaring event, its orbit is distorted during the flux eruption stage where the inner layers associated with the  negative X values experience an outflowing motion, while at further distances the flow remains clockwise.

\section{Conclusion}
\label{Conclusion}
In this study, we performed a comprehensive theoretical analysis of a flaring event in M87*, leveraging a suite of GRMHD simulations to explore the underlying physics. Flares were identified as the peaks in flux intensity over a simulated 15-year observational period, with variations in BH spin and two distinct magnetic field configurations: magnetically arrested disk (MAD) and the weakly magnetized case, known as the standard and normal evolution (SANE). Our analysis revealed that the flare scenario in the MAD configuration with a BH spin of a=+0.5 is the most physically plausible. Consequently, we concentrated our detailed case study on this configuration. Below, we outline the key insights derived from our investigation: \\

$\bullet$ As illustrated in Figure \ref{Image-Flare-ap5}, the flux flare is succeeded by a flux eruption, during which the flux density generates an outflowing current that propagates to larger distances. We argue that it may mean that a millimeter/radio flare probably leads to a flare in Near infrared and the X-ray bands as well. As an evidence in SgrA*, the flux eruption is related to the X-ray and NIR flares. 
For Sgr A*, flux eruptions have been related to Xrays and nIR flares \citep{2021ApJ...919L..20D, 2021MNRAS.502.2023P,2022ApJ...924L..32R}.

$\bullet$ As the flux propagates outward, the inferred emission location, as shown in Figure \ref{Emission-location}, expands from the equatorial plane to regions beyond the Z=0 surface. 

$\bullet$ As shown in Figure \ref{Stokes_Flare}, both linear and circular polarization exhibit pronounced peaks in the vicinity of the flaring event. Additionally, the  EVPA undergoes a sign reversal near the flare.

$\bullet$ An analysis of different $\beta_m$ modes in the vicinity of the flux flare reveals, as shown in Figure \ref{Betam_flare}, that $\beta_2$ exhibits no significant sensitivity to the flare. In contrast, other modes display more pronounced variations during the flux flare.

$\bullet$ Figure \ref{Image_Flare_vis} illustrates the map of the logarithm of the visibility amplitude, demonstrating that in the vicinity of the flux flare, the map is nearly symmetric and round. However, its morphology evolves before and after the flare, adopting a more elliptical shape.

$\bullet$ Figure \ref{Flare_Vis_Baseline} indicates that intermediate baselines exhibit greater sensitivity to the flare compared to short or long baselines.

$\bullet$ Figure \ref{Image_Flare_EB_phase} clearly demonstrates that the phase of the 
$EB$-correlation is highly sensitive to various stages of the flare, including the flux growth, onset, and main phase of the flux eruption. Consequently, we conclude that the 
$EB$-correlation phase is the most sensitive metric for capturing transient variability in the vicinity of the flux flare.

$\bullet$ Figure \ref{light-curve-subring-fig} reveals that photon sub-rings closely mirror the overall flux variations, exhibiting a strong correlation with both the increasing and decreasing phases of the flux.

$\bullet$ The maps of the visibility amplitude and the $EB$-correlation phase, as shown in Figures \ref{Visibility-subring} and \ref{EB-correlation-subring}, are analyzed for all-ns, ns=0, and ns=1. The results indicate that while the morphological structures of the visibility amplitude and correlation phase are closely aligned for the all-ns and ns=0 cases, they differ significantly for ns=1. This distinction underscores the potential of future observations of the ns=1 component in the vicinity of intensity flares to provide unique and invaluable insights distinct from those offered by other image moments.

$\bullet$ Depictions of the magnetic field (Figures \ref{B-fieldXZ}-\ref{B-fieldXY}) in the vicinity of the flaring event reveal that magnetic reconnection and loops diminish during the flux eruption phase. Similarly, the velocity field maps (Figures \ref{V-fieldXZ}-\ref{V-fieldXY}) indicate that the clockwise motion of the flow transitions into an outflow during the flux eruption phase.

\section{Future direction}
\label{future}
While this work primarily focuses on an in-depth analysis of structural variations for the case of a BH spin of $a=+0.5$ with an isolated flare, future studies will extend this investigation to the case of $a=-0.93$, where flares occur more frequently. A comprehensive comparison of structural variations in the vicinity of flaring events may provide valuable insights into the nature of BH spin. We leave this exploration for future work.

While this work primarily focuses on $R_{\mathrm{low}} = 1$, $R_{\mathrm{high}} = 20$, and $\sigma_{\mathrm{cut}} = 1$, future analyses will extend this framework  to explore different combinations of ($R_{\mathrm{low}}$, $R_{\mathrm{high}}$) \citep{2023MNRAS.526.2924J} and variations in the value of $\sigma_{\mathrm{cut}}$.

Finally, although the current suite of GRMHD simulations is non-radiative, it would be highly compelling to extend this investigation to radiative GRMHD simulations. We leave such an exploration to future work.

\section*{Data Availability}
The data associated with this work and its figures will be made available upon reasonable request to the corresponding author. The ray tracing employed in this study was conducted using the {\sc ipole} method \citep{Moscibrodzka&Gammie2018}. We utilized 3D time-dependent GRMHD simulations made using the H-AMR code \citep{2022ApJS..263...26L}. These simulations are part of the GRMHD simulation set described in \citet{2022ApJ...930L..16E}, and in \citet{Chatterjee:2022}.

\section*{Acknowledgment}
We are deeply grateful to Shep Doeleman, Daniel Eisenstein, Ramesh Narayan, and George Wong for insightful discussions and valuable contributions. Razieh Emami warmly acknowledges the generous support of the Institute for Theory and Computation (ITC) at the Center for Astrophysics. We also thank the supercomputer facility at Harvard University, where the majority of the simulation work was conducted. 
An award of computer time was provided by the Innovative and Novel Computational Impact on Theory and Experiment (INCITE) and ASCR Leadership Computing Challenge (ALCC) programs under award AST178. This research used resources of the Oak Ridge Leadership Computing Facility, which is a DOE Office of Science User Facility supported under Contract DE-AC05-00OR22725. ML was supported by the John Harvard, ITC and NASA Hubble Fellowship Program fellowships, and NASA ATP award 21-ATP21-0077. MW is supported by a~Ramón y Cajal grant RYC2023-042988-I from the Spanish Ministry of Science and Innovation.

\textit{Software:} matplotlib \citep{2007CSE.....9...90H}, numpy \citep{2011CSE....13b..22V}, scipy \citep{2007CSE.....9c..10O}, seaborn \citep{2020zndo...3629446W}, pandas \citep{2021zndo...5203279R}, h5py \citep{2016arXiv160804904D}.

\appendix 
\section{Damped Random Walk}
\label{DRW_Details}
Here we present the details of the DRW fitting procedure. While the original DRW was presented by \cite{2009ApJ...698..895K}, 
we follow the modified DRW process by   \citep{2022ApJ...930L..19W} by including an extra parameter $\sigma_0$ as an indication of the noise floor. Consequently the DRW is described with four parameters as:

\begin{equation}
\theta = (\tau,
\mu, \sigma, \sigma_0).
\end{equation}
The likelihood function for observations, $ \lbrace x_i  \rbrace = x_1, x_2, ..., x_n$ at the times $ \lbrace t_i \rbrace = t_1, t_2, ..., t_n$ is given by:

\begin{equation}
\label{Likelihood-DRW}
\mathcal{L} = \prod_{i =1}^{n} \frac{\exp{(-0.5(\hat{x}_i - x_i^*)^2/\widetilde{\Omega}_i)}}{(2\pi \widetilde{\Omega}_i)^{1/2}}, 
\end{equation}
where,
\begin{equation}
\widetilde{\Omega}_i =  \Omega_i + \sigma^2_0 + \sigma^2_i, ~~~~~~~~~
x_i^* =  x_i - \mu.
\end{equation}
The quantities $\Omega_i$ and $\hat{x}_i$ are computed through an iterative process:
\begin{equation}
\label{xi}
\begin{split}
\hat{x}_i = & a_i \hat{x}_{i-1} + \frac{a_i \Omega_{i-1}}{\widetilde{\Omega}_{i-1}}(x^{*}_{i-1} - \hat{x}_{i-1}),  \\ 
\Omega_{i} = & \Omega_1 (1 - a^2_i) + a^2_i \Omega_{i-1} \left( 1 - \frac{\Omega_{i-1}}{\widetilde{\Omega}_{i-1}} \right) \\
a_i = & \exp{(-\Delta t_i/\tau)} 
\end{split}
\end{equation}
where the initial conditions are given by: 
\begin{equation}
\Omega_1 = \sigma^2, ~~~~~~ \hat{x}_i = 0. 
\end{equation}
We use a Bayesian approach with the following priors for the DRW parameters: 

\begin{equation}
\begin{split}
\log_{10}{(\tau/M)} = & \mathcal{N}_T(0.5, 1), \\
\sigma (Jy) = & \mathcal{N}_T(0.1, 1), \\
\log_{10}{(\mu/Jy)} = & \mathcal{N}_T(-0.3, 1), \\
\sigma_0 (Jy) = & \mathcal{U}(0.0, 0.1).
\end{split}
\end{equation}
where $\mathcal{N}_T(a,b)$ and $\mathcal{U}(a,b)$ refer to the normal and uniform distribution with the mean $a$ and the standard deviation $b$, respectively.

\bibliography{main}

\end{document}